\PassOptionsToPackage{square,comma,numbers,sort&compress,super}{natbib}
\documentclass[manuscript]{aastex62}
\usepackage{amsmath}
\usepackage{gensymb}
\usepackage{graphics}
\usepackage{subfigure}
\usepackage{multirow}
\usepackage{enumerate}
\usepackage{longtable,booktabs}

\submitjournal{ApJ}

\shorttitle{SFR Observations at Ulysses}
\shortauthors{Chen et al.}

\begin{document}


\correspondingauthor{Qiang Hu}
\email{qh0001@uah.edu}

\author[0000-0002-0065-7622]{Yu Chen}
\affiliation{Department of Space Science, The University of Alabama in Huntsville, Huntsville, AL 35805, USA}

\author[0000-0002-7570-2301]{Qiang Hu}
\affiliation{Department of Space Science, The University of Alabama in Huntsville, Huntsville, AL 35805, USA}
\affiliation{Center for Space Plasma and Aeronomic Research (CSPAR), The University of Alabama in Huntsville, Huntsville, AL 35805, USA}

\author[0000-0001-9199-2890]{Jakobus A. le Roux}
\affiliation{Department of Space Science, The University of Alabama in Huntsville, Huntsville, AL 35805, USA}
\affiliation{Center for Space Plasma and Aeronomic Research (CSPAR), The University of Alabama in Huntsville, Huntsville, AL 35805, USA}

\begin{abstract}

Small-scale magnetic flux ropes, in the solar wind, have been
studied for decades via the approach of both simulation and
observation. Statistical analysis utilizing various in-situ
spacecraft measurements is the main observational approach. In
this study, we extend the automated detection of small-scale flux
ropes based on the Grad-Shafranov reconstruction to the complete
dataset of \emph{Ulysses} spacecraft in-situ measurements. We
first discuss the temporal variation of the bulk properties of
22,719 flux ropes found through our approach, namely, the average
magnetic field and plasma parameters, etc., as functions of the
heliographical latitudes and heliocentric radial distances. We
then categorize all identified events into three groups based on
event distributions in different latitudes separated at
30$^{\circ}$, at different radial distances, and under different
solar activities, respectively. By the detailed statistical
analysis, we conclude as follows. (1) The properties of flux
ropes, such as the duration, scale size, etc., follow the
power-law distributions, but with different slope indices,
especially for distributions at different radial distances. (2)
Also, they are affected by the solar wind speed which has
different distributions under the different solar activities,
which is manifested as the latitudinal effect. (3) The main
difference in flux rope properties between the low and high
latitudes is attributed to possible Alfv\'enic structures or
waves. (4) Flux ropes with longer duration and larger scale sizes
occur more often at larger radial distances. (5) With more strict
Wal\'en slope threshold, more events are excluded at higher
latitudes. The entire database is published online at
\url{http://www.fluxrope.info}.
\end{abstract}

\title{Analysis of Small-scale Magnetic Flux Ropes Covering the Whole \emph{Ulysses} Mission}
\section{Introduction}
A small-scale magnetic flux rope (hereafter, SFR), introduced by \cite{Moldwin1995,Moldwin2000}, is defined by a magnetic configuration with helical and winding magnetic field lines, the same as the large-scale flux rope but with much smaller scale size and duration range. Observed from in-situ spacecraft measurements, e.g., at 1 au, this type of structure always has the following characteristics: (1) The magnetic field components from time series data have twisting and continuous rotation. (2) The structure is in quasi-static equilibrium and convected with the solar wind.

It is still controversial where SFRs originate. Some authors suggest that this small-scale structure has features in common with its large-scale counterpart, the magnetic cloud. Based on investigation of the counterstreaming suprathermal electron signatures and distributions of SFR characteristics via \emph{Wind} spacecraft measurements, they believe that these small-scale structures may come from solar eruptions and are manifestations of small coronal mass ejections \citep{Feng2007, Feng2008}. However, there is some doubt about their solar origins since their occurrence rate has weak solar cycle dependency. Instead, magnetic reconnection across the heliospheric current sheet (HCS) is considered to be a possible source \citep{Cartwright2008}. Later, \cite{Tian2010} pointed out that both the Sun and processes in interplanetary space can generate SFRs. Through STEREO observations, \cite{Yu2016} discussed the characteristics of small solar transients (STs), structures including SFRs, and suggested that STs have opposite solar cycle dependency. Furthermore, two sources, the solar corona and the interplanetary medium, are verified to be factors affecting the occurrence of STs.

On the basis of available data from multiple spacecraft measurements, such as \emph{ACE}, \emph{Wind}, \emph{Helios} and \emph{Ulysses} for the past few years, statistical investigation of small-scale flux ropes becomes essential for the determination of their origin and obtaining a better understanding of their evolution. In many analysis techniques, Grad-Shafranov (GS) reconstruction is the outstanding and efficient one which can recover two-dimensional (2D) structures from one-dimensional (1D) in-situ spacecraft measurements. This method was first introduced by \cite{Sonnerup1996,Hau1999} and then applied to reconstructing magnetic flux ropes in the solar wind by \cite{Hu2001, Hu2002}. Recently, a series of research on SFRs publications appeared in which the GS reconstruction technique was used to analyze SFRs at 1 au. \cite{Zheng2018} identified 74,241 SFRs via \emph{Wind} spacecraft measurements by designing and carrying out the automated detection based on the GS reconstruction. With this event database, they showed that these structures have a visible solar cycle dependency and that the probability density function of the axial current density distribution has non-Gaussian features in correspondence with simulation results. The detailed algorithm and statistical analysis are described further in \cite{Hu2018}. In this report, they showed that many properties including both duration and scale sizes of SFRs obey power laws, and that many SFRs are often accompanied by the HCS crossing within a day. Furthermore, the cycle-to-cycle variation of major SFRs statistics was also examined, which yields little difference.

Recently, this automated detection has been applied to \emph{ACE} spacecraft with the full SFR lists and selected plots supplied on the website of the small-scale magnetic flux rope database \url{(http://www.fluxrope.info)}. Since the present two databases are limited to spacecraft missions around 1 au and near ecliptic plane, the next step is to explore the characteristics of SFRs at different radial distances and helio-latitudes.

The \emph{Ulysses} spacecraft is the best choice for this task since it provides the in-situ measurements covering a wide range of heliographic latitudes and radial distances in the solar wind. The spacecraft was launched in October 1990 with the primary mission to probe the solar poles and almost all latitudes around the Sun. In each orbit, it reached the farthest distance (5.41 au), and the north as well as the south poles (around +80.2$^{\circ}$ to -80.2$^{\circ}$ latitude). With multiple instruments onboard, it detected solar wind plasma, interplanetary magnetic field, solar and galactic cosmic rays, etc. In particular, the Vector Helium Magnetometer (VHM) instrument \citep{Balogh1992} provides the magnetic field data including three components in the RTN (radial, tangential and normal) coordinate system and field magnitude. The Solar Wind Observations Over the Poles of the Sun (SWOOPS) instrument \citep{Bame1992} detects the plasma parameters, such as the ion density, the solar wind flow velocity, which is also in the RTN coordinates, and the proton temperature ($T_{large}$ and $T_{small}$). The detailed distribution of solar wind speed over the first two complete \emph{Ulysses} orbits was presented in \cite{McComas2003}. They showed that the first orbit began when the Sunspot number was in declining phase while the second orbit coincided with the solar maximum period. These observations presented in two polar plots depicted the classical picture that during solar minimum period, the high speed wind occurs mainly at higher latitudes, well separated from low speed wind at lower latitudes, while during solar maximum, high speed and low speed wind streams are intermixed. Finally, the \emph{Ulysses} spacecraft was switched off in the middle of 2009 after accomplishing a 18.5-year long mission and three complete orbits \citep{McComas2013}.

In a previous study \citep{Chen2018}, we reported the SFR detection results from \emph{Ulysses} for four specific years (1994, 1996, 2004 and 2005) during the solar minimum periods. These four years are categorized into two groups, i.e., one with high speed wind or high latitudes and the other with low speed wind or low latitudes. Despite the limited number of recorded events, most of the properties of SFRs still exhibit power-law distributions and non-Gaussian features which are consistent with analysis results at 1 au. This set of SFRs was also used to assist in other relevant studies. For example, unusual energetic particle flux enhancements in February 2004 were discussed in \cite{Zhao2018, Zhao2019} in association with the emerging SFRs identified in our event set.

In this study, we further extend the previous event set by applying our automated detection technique to \emph{Ulysses} spacecraft measurements covering the full 18.5 years mission. The observational analysis of the identified small-scale flux ropes will be presented in the following order. An introduction to Grad-Shafranov reconstruction and new criteria for the automated detection are described in Section \ref{sec:gs}. An overview of the principal characteristics of SFRs, such as the magnitude of magnetic field, solar wind speed, plasma parameters together with their yearly variations are presented in Section \ref{sec:overview}. Due to the unique orbits of \emph{Ulysses}, we separate the SFR dependence on latitude as well as radial distance and discuss the properties of SFRs under different circumstances in Sections \ref{sec:lat} and \ref{sec:rad}. Section \ref{sec:solar} demonstrates the features of these structures for different solar activity levels. The conclusions from our analysis and the applications of our existing and future databases are discussed in the last section.

\section{Grad-Shafranov Reconstruction and the Automated Detection Algorithm}\label{sec:gs}
The new database of small-scale magnetic flux ropes for \emph{Ulysses} in this study is obtained via the automated detection algorithm, which was introduced in \cite{Zheng2018}, based on the technique of the Grad-Shafranov (GS) reconstruction. The detailed study based on \emph{Wind} spacecraft data and a flowchart of the algorithm are presented in \cite{Hu2018} to illustrate the procedures for computer inplementation (see also www.fluxrope.info).

The standard GS equation is given by:
\begin{equation}
\frac{\partial^2{A}}{\partial{x}^2}+\frac{\partial^2{A}}{\partial{y}^2}=-\mu_0\frac{dP_t(A)}{dA}=-\mu_0j_z(A),
\end{equation}

where the transverse pressure ${P_t}$ is defined as $P_t=p+({B_z}^2/{2\mu_0})$, the sum of the plasma pressure $p$ and the axial magnetic field pressure, both of which are single-variable functions of the magnetic flux function $A(x, y)$. The transverse magnetic field components are given by $B_x = \partial A/\partial y$, and $B_y = -\partial A/\partial x$, respectively. Therefore, the cross section of each flux rope in the $(x, y)$ plane, described by $A$, the axial field $B_z$ and the axial current density $j_z$, can be determined by the solution to the GS equation. As a unique feature, the GS reconstruction recovers 2D cross section from 1D in-situ spacecraft measurements. When discussing in-situ detection of a flux rope, it is critical to realize that a spacecraft will generally collect data across an equivalent set of helical magnetic field lines when it crosses from one edge to the center of each flux rope, then from the center to the other edge (in reverse order) along the $x$-axis, i.e., $y = 0$. The point when the sign of the field component, ${B_y}$, begins to switch is marked as the turning point. At the turning point, the magnetic flux function $A$ reaches its maximum or minimum along $y = 0$. Thus, the spacecraft path crossing a flux rope can be split into two branches at the turning point. One branch will fold back onto the other in terms of their corresponding $A$ values, whose range in absolute magnitude usually lies between $\sim$ 0 and the absolute extremum, corresponding to the boundary and the center of the flux rope, respectively. In view of any quantities being single-variable functions of $A$, these two branches, for example, as represented by the $P_t$ versus $A$ curves, shall overlap and it is regarded to double-folding pattern.

\begin{table}
\begin{center}
\caption{The criteria of small-scale magnetic flux rope detection for \emph{Ulysses}.}
\begin{tabular}{ccccc}
\hline
Duration (minutes) & $R_{dif}$ & $R_{fit}$ & $Wal\acute{e}n~test~slope$\\
\hline
$45\sim2255$ & $\le{0.12}$ & $\le{0.14}$ & $\le{0.5}$\\
\hline
\end{tabular}
\label{table:table1}
\end{center}
\end{table}

With this particular one-to-one correspondence and double-folding behavior, the automated detection algorithm is built. The new set of search criteria is applied to the \emph{Ulysses} dataset (Table \ref{table:table1}). First, we set up the duration range of search windows. In the previous study, \cite{Zheng2018} used 9 $\sim$ 361 minutes for the \emph{Wind} spacecraft database of SFRs. Here we modify it to 45 $\sim$ 2,255 minutes for \emph{Ulysses}. The main reason for setting longer duration limits is that the plasma parameters required to calculate the thermal pressure and frame velocity are with 4 to 8 minutes cadence, and we extend the upper limit to $\sim$ 37 hours in order to accommodate relatively large-scale flux ropes or magnetic clouds. Starting at the first minute of each year, multiple sliding search windows, with sizes ranging from (45, 80), (70, 105), $\dots$, and (2145, 2255) minutes, are adopted. These duration ranges enable our results to cover all SFRs of variable sizes as applied to the entire \emph{Ulysses} dataset. Among these duration ranges, the second number is the maximum length of interval or data array for the calculation of parameters, e.g., ${P_t}$, whereas the first number is the lower limit of the length for finding double-folding behavior of ${P_t}$ versus $A$. Additionally, there is 10 minutes overlapping between adjacent search windows in order to preserve candidate events as many as possible, and ensure a smooth transition from one search window to the next \citep{Hu2018}.

Next, a quasi-stationary frame of reference is necessary for all subsequent calculations since the GS equation is derived assuming a 2D magnetohydrostatic equilibrium. Thus, all \emph{Ulysses} data including magnetic field components and solar wind velocity are transformed to the new frame, i.e., the deHoffmann-Teller (HT) frame, generally. This new frame is obtained by the determination of the HT frame velocity, ${\bf{V}}_{HT}$, which can be computed by the solar wind and the magnetic field data \citep{Hu2002}. Sometimes, the average solar wind velocity is utilized as ${\bf{V}}_{HT}$ to ensure the efficiency of calculation. Then, we follow the basic steps of the Grad-Shafranov reconstruction by \cite{Hu2002} and perform minimum-variance analysis on the measured magnetic field (MVAB) to find the trial frame for the GS reconstruction, in which the validity of the underlying 2D geometry can be checked, by varying the trial $z$-axis and calculating the corresponding metric (residue; defined below) assessing the satisfaction that $P_t$ be a function of $A$. In such a frame, including the one with the optimal $z$-axis with the minimum residue, the $x$-axis is projected along the spacecraft path, and $y$-axis is determined following the right-handed orthogonal coordinate system.

In this frame, the calculation of the magnetic flux function $A(x, 0)$ and the transverse pressure $P_t$ is carried out along the spacecraft path, $y = 0$. With the initial data $B_y(x,0)$ from the spacecraft measurements, the quantity $A(x, 0)$ is obtained by

\begin{equation}
A(x,0)=\int_{0}^{x}\frac{\partial{A}}{\partial\xi}d\xi=\int_{0}^{x}-B_y(\xi,0)d\xi,
\end{equation}
with $d\xi=-{\bf{V}}_{HT}\cdot{\bf\hat{x}}dt$, where $dt$ is the time increment. The calculation of $P_t$ will be executed if there is only one turning point of $A$ within this time interval. As referred above, the transverse pressure $P_t$ consists of the thermal pressure $p=N_pkT_p$ the product of proton number density $N_p$, the Boltzmann constant $k$, and the proton temperature $T_p$ (the electron temperature $T_e$ is unavailable from \emph{Ulysses}, and $T_p = T_{small}$ is adopted as the proton temperature), and the axial magnetic pressure ${B_z}^2/{2\mu_0}$, which is also known from measurements.

As emphasized earlier, there exists the double-folding behavior in $P_t$ versus $A$ curves for a flux rope solution, which is a predominant criterion for this study. Evaluation of this behavior is initiated by finding the turning point of $A$ where the sign change of the field component $B_y$ and the corresponding split of the $P_t$ versus $A$ curve into two overlapping branches occur. Once this turning point is verified, together with all the calculations of $P_t$ in these search windows, two residues, the difference residue and the fitting residue, are obtained \citep{Hu2002, Hu2004, Hu2018}, as follows:
\begin{equation}
R_{dif}=\frac{1}{2}[\frac{1}{N}\sum_{i=1}^{N}((P_t)_i^{1st}-(P_t)_i^{2nd})^2]^{\frac{1}{2}}/|max(P_t)-min(P_t)|,
\end{equation}
and
\begin{equation}
R_{fit}=[\frac{1}{L}\sum_{i=1}^{L}(P_t(x_i,0)-P_t(A(x_i,0)))^2]^{\frac{1}{2}}/|max(P_t)-min(P_t)|.
\end{equation}

These two residues indicate the quality of the double-folding pattern of ${P_t}$ versus $A$ curves. As shown in the above equations, with all $P_t$ values of each branch (denoted by ``1st'' and ``2nd'') within the flux rope interval, the average difference between these two branches is calculated and normalized by the difference between the minimum and the maximum ${P_t}$. Both $L$ and $N$ represent the total length of each data array involved. As the $R_{dif}$ is evaluated between two branches, while the other residue $R_{fit}$ is evaluated with respect to the fitting ruction $P_t(A)$,
the additional fractional factor $1/2$ for $R_{dif}$ is added to approximately account for the the discrepancy in their average magnitudes. In this study, the thresholds are set as 0.12 and 0.14, respectively, based on our experience.

Furthermore, an extra criterion is crucial to ensure the validity of quasi-static equilibrium underlying the GS equation. The Wal\'en relation in the HT frame, introduced by \cite{Paschmann1998}, is used to evaluate the ratio of the remaining flow velocity to the local Alfv\'en velocity. In other words, a small Wal\'en slope threshold will exclude Alfv\'enic structures and waves which do not fall into the categorization as SFRs, governed by the GS equation. Last but not the least, the clean-up process is implemented. In this final step, candidate events with overlapping time intervals will be filtered by a combined approach of sorting the turning points and the minimized residues to ensure a list of identified events with distinct intervals (see \cite{Hu2018} for details).

\section{Overview of Small-scale Flux Rope Detection Results at \emph{Ulysses}}\label{sec:overview}
The automated detection is completed, following the approach of \cite{Hu2018} as described in the previous section, covering the whole \emph{Ulysses} mission from 1991 to the middle of 2009. For these 18.5 years, the total number of small-scale flux ropes  detected is 22,279. We provide an overview of the detection results in the following three aspects: the change of SFRs' monthly-averaged bulk properties, the SFRs monthly counts and the detailed SFR categories, in order to facilitate the subsequent statistical analysis.

\subsection{Temporal Change of Small-scale Flux Rope Bulk Properties}
\begin{figure}
\begin{centering}
\includegraphics[scale=0.62]{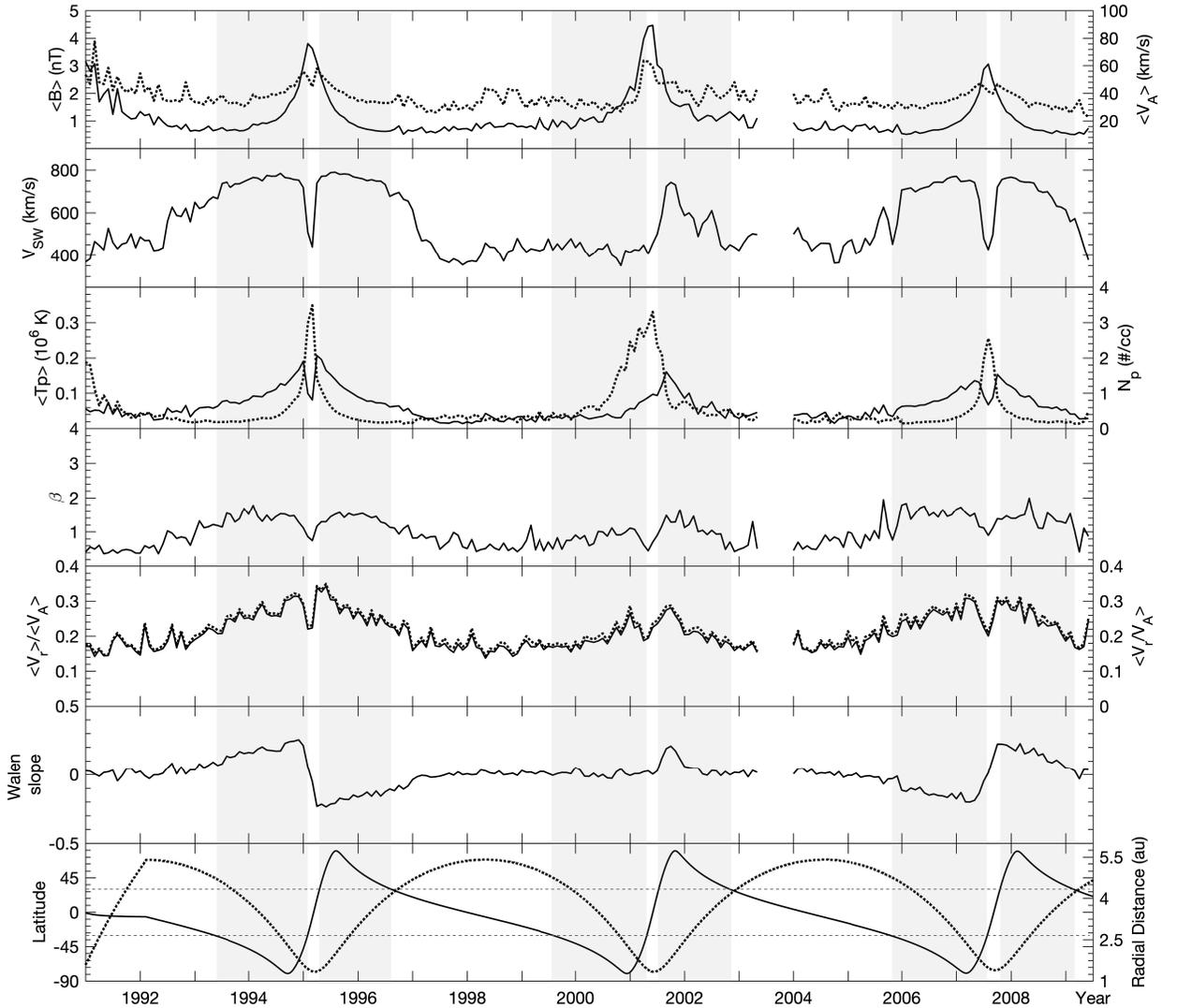}
\caption{Monthly averages of small-scale flux ropes' characteristics from 1991 to the middle of 2009, spanning the whole \emph{Ulysses} mission. From the top to the bottom panels are: (1) The average magnetic field magnitude and average Alfv\'en speed. (2) The average solar wind speed. (3) The average proton temperature $T_{small}$ and number density $N_p$. (4) The plasma proton $\beta$, i.e., the ratio of thermal pressure to magnetic pressure. (5) The Alfv\'en Mach number, defined as the ratio of the remaining flow speed to the Alfv\'en speed. The left $y$-axis takes average calculation before determining the Mach number while the right $y$-axis is its average value. (6) Wal\'en slope. (7) Hourly merged orbit data in terms of the heliocentric distance and Heliographic Inertial (HGI) latitude for the whole \emph{Ulysses} mission. All average values are calculated within each flux rope interval and averaged over one month. The solid curves correspond to quantities with labels and scales given by the left axis, while the dotted curves have labels and scales given by the right axis. The grey shaded areas denote the \emph{Ulysses} orbital periods with the HGI latitude beyond $\pm$ 30$^{\circ}$.}\label{fig:overview}
\end{centering}
\end{figure}

The temporal change of the principal characteristics of small-scale flux ropes is presented in Figure \ref{fig:overview}. The magnetic field magnitude of SFRs, plotted in the first panel, shows that the variation is associated with both radial distance and latitude. The field magnitude peaks when \emph{Ulysses} went to the smallest radial distance or can be ascribed to decreasing latitudes. However, this tendency caused by the latter factor if any does not show up when the latitude changes gradually, e.g., for years 1997 - 2000 when the spacecraft roamed to far radial distances recording extended period of weak magnetic field (less than 1 nT on average). Hence, the variation of field magnitude of SFR is related mostly to change of radial distances.

The plasma parameters, including solar wind speed, proton temperature, number density and plasma $\beta$ are plotted in the 2nd, 3rd and 4th panels, respectively. Most of high speed solar wind occurs at relatively high latitudes during the solar minimum periods. Notice that there are two troughs surrounded by plateaus. They represent low latitude regions in which \emph{Ulysses} spent two to three months each passing through when the rapid pole to pole transition happened, although such a pattern was much disrupted during the maximum years around 2001. During the solar maximum, fast solar wind appears to occur more irregularly at all latitudes, as reflected in our detection results with flux ropes of variable solar wind speed not clearly separated in latitudes, in contrast to those during solar minimum.

The proton number density increases substantially during those fast latitude scans (sandwiched between each pair of closely spaced shaded areas in Figure \ref{fig:overview}) where the radial distances are about 1 $\sim$ 2 au. This is an indication that the solar wind plasmas are denser close to the Sun. In addition, the density does not always follow the change of solar wind speed, especially during the period of maximum. The average proton temperature, on the contrary, does not have such corresponding changes with radial distances but seems to correlate better with solar wind speed. In particular, the temperature peaks with solar wind speed enhancement during the solar maximum (the second half of year 2001). As a combined effect, the large field magnitudes, sudden changes of proton number density and low proton temperature at these small radial distances produce localized troughs in plasma $\beta$, but the average Alfv\'en speed still peaks broadly at times of maximum $<B>$, whereas the plasma ${\beta}$ dips accordingly.

Both the Alfv\'en Mach numbers (5th panel) and Wal\'en slope values (6th panel) represent the ratio between the remaining flow speed to the local Alfv\'en speed. Despite that these two parameters fluctuate within a narrow range for most of the time, they increase in magnitude when the \emph{Ulysses} approached the latitudinal extrema in both northern and southern hemispheres. This trend is especially well preserved for solar minimum years, but much diminished during the solar maximum. Those latitudes are the possible places where the Alfv\'enic structures or waves are prevailing, resulting in relatively large Wal\'en slopes, for instance. The existence of these structures will be further explored in the next subsection.

\subsection{Temporal Change of Small-scale Flux Rope Occurrence Rate}
\begin{figure}
\begin{centering}
\includegraphics[scale=0.55]{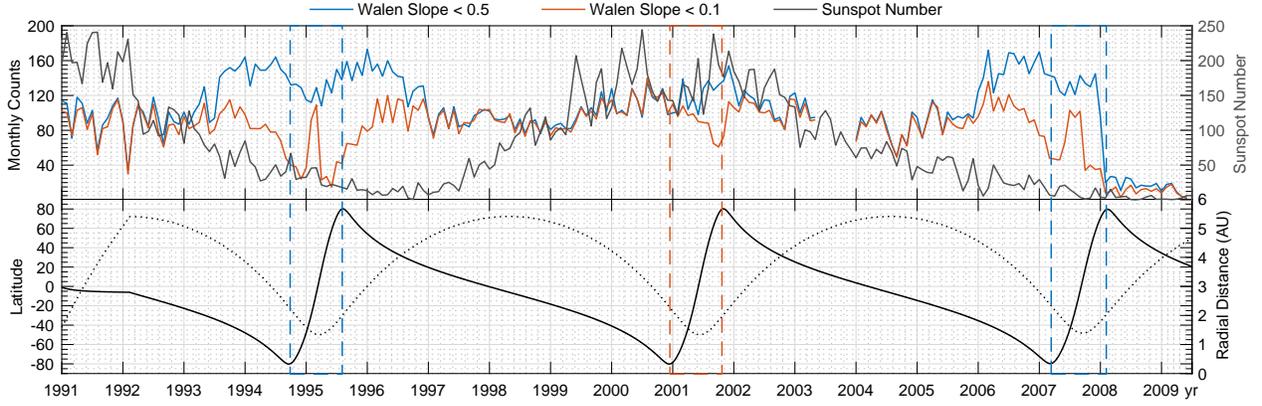}
\caption{The monthly counts of flux ropes with yearly change. The top panel shows monthly counts with different Wal\'en slope threshold (0.5 and 0.1) and monthly average sunspot number from Sunspot Index and Long-term Solar Observations (SILSO), marked by blue, dark orange and grey curves, respectively. The bottom panel shows the HGI latitudes and radial distances of the \emph{Ulysses} spacecraft orbit around the Sun. The pairs of blue and dark orange dashed lines enclose the three fast latitudinal scans of \emph{Ulysses} orbit when the solar activity nearly reached its minimum and maximum levels. }\label{fig:ulycount}
\end{centering}
\end{figure}

The connection between the solar activity and the properties of small-scale flux ropes is what we are curious about the most. Similar to the \emph{ACE} or \emph{Wind} spacecraft, the \emph{Ulysses} mission covers the whole solar cycle 23, and the declining phase of cycle 22 as well. Figure \ref{fig:ulycount} shows the monthly occurrence rate of flux ropes filtered by different Wal\'en slope thresholds (0.5 and 0.1, respectively) and the change of the monthly sunspot number for almost 1.5 solar cycles. With 0.5 as the threshold, there are three broad peaks in monthly counts residing at the corresponding solar minimum, maximum and minimum, respectively. During the solar maximum around year 2001, we notice that the monthly counts follow the number of sunspots well. The average monthly counts are approximately the same at different latitudes, for approximately the same radial distances within a narrow range. This consistency is probably due to non-differentiating distribution of solar wind in latitude at solar maxima. During the solar minima, the peaks in blue generally correspond to high latitudes where high speed wind dominates. For the low latitudes (0 $\sim$ 30$^{\circ}$), more flux ropes are detected at smaller radial distances.

Considering the possible Alfv\'enic structures and waves at high latitudes as suggested in Figure \ref{fig:overview}, we set the more strict criterion, i.e., 0.1 as Wal\'en slope threshold. \cite{Marubashi2010} suggested that torsional Alfv\'en waves are actually pseudo flux ropes which may be identified mistakenly as flux ropes (see also \cite{Higginson2018}). In the same year, \cite{Gosling2010} described a rare case of torsional Alfv\'en waves embedded within small-scale flux ropes. Therefore, this more strict limit not only excludes real Alfv\'enic structures or waves, but also flux ropes with relatively high Alfv\'enicity. The total number of flux ropes under this new threshold is 17,660 with a 22.2\% reduction. In this percentage, 19.4\% is from relatively high latitudes (greater than or equal to 30$^{\circ}$) while only 2.8\% is from latitudes less than 30$^{\circ}$. Such contribution to this reduction is also anticipated by Figure \ref{fig:overview} which shows that most of large Wal\'en slope values arise at high latitudes. With this new set of detection results, three peaks in Figure \ref{fig:ulycount} (top panel) no longer exist. Instead, although with fluctuations, the occurrence rate does not have prominent peaks, but appears flat with rather sudden drops at high latitudes especially during solar minima. Note that near 0$^{\circ}$ latitude, the monthly counts are nearly unchanged from the result with 0.5 as the Wal\'en slope threshold. The occurrence rate drops abruptly in year 2008 under both thresholds. This is because of large data gaps in magnetic field and plasma parameters while our detection is based on the availability of 1 minute resolution data.

In summary, we may be expecting to obtain a similar solar cycle dependency of SFRs at \emph{Ulysses} occurrence as revealed by 1 au detection results, but apparently, the monthly counts of SFR are modulated by the varying radial distances and latitudes of the unique \emph{Ulysses} orbits which may conceal or overrun the effects of solar activity. The detailed comparison of results under different solar activity levels will be discussed in Section \ref{sec:solar}.

\subsection{Categorization of Small-scale Flux Rope Database}
\begin{table}
\begin{center}
\caption{Category of small-scale magnetic flux ropes.}
\footnotesize
\begin{tabular}{ccccccc}
\hline
Category & & Time Periods & Latitudes & Radial Distances & Event Counts\\
& & (YYYY-mm-dd) & ($^{\circ}$) & (au) & \\
\hline
\multirow{12}*{I. Latitude} & & 1991-01-01 $\sim$ 1993-05-02 & & \\
& & 1995-01-23 $\sim$ 1995-04-12 & & \\
& & 1996-08-02 $\sim$ 1999-07-14& $<$ 30 & 1.34 $\sim$ 5.41 & 9,595 \\
& & 2001-04-06 $\sim$ 2001-06-23& & & (8,950\tablenotemark{a}) \\
& & 2002-10-12 $\sim$ 2005-09-14& \\
& & 2007-06-29 $\sim$ 2007-09-21& \\
\cline{3-6}
& & 1993-05-02 $\sim$ 1995-01-23 & \\
& & 1995-04-12 $\sim$ 1996-08-02 & \\
& & 1999-07-14 $\sim$ 2001-04-06 & $\geqslant$ 30 & 1.34 $\sim$ 5.41 & 13,124 \\
& & 2001-06-23 $\sim$ 2002-10-12 & & & (8,711\tablenotemark{a})\\
& & 2005-09-14 $\sim$ 2007-06-29 & \\
& & 2009-01-22 $\sim$ 2009-06-30 & \\
\hline
\multirow{3}*{II. Radial Distances} & & 1998-02-05 $\sim$ 1999-07-14 & $\sim$ 0 & $\sim$ 1 au & 17,620\tablenotemark{b} \\
& & 2002-10-12 $\sim$ 2005-09-14 & \\
\cline{3-6}
& & 1998-02-05 $\sim$ 1999-07-14 & $<$ 30 & $>$ 3.5 au & 10,124 \\
& & 2002-10-12 $\sim$ 2005-09-14 & \\
\hline
\multirow{3}*{III. Solar Activity} & Maximum & 2000-11-22 $\sim$ 2001-10-13 & -80.2 $\sim$ 80.2 & 1.34 $\sim$ 2.31 & 1,238 \\
\cline{2-6}
& Minimum & 1994-09-10 $\sim$ 1995-08-02 & -80.2 $\sim$ 80.2 & 1.34 $\sim$ 2.32 & 2,912 \\
& & 2007-02-04 $\sim$ 2008-01-17 & -79.7 $\sim$ 79.7 & 1.39 $\sim$ 2.41 & \\
\hline \noalign {\smallskip}
\end{tabular}
\tablenotetext{a}{Events detected under Wal\'en slope threshold 0.1.}
\tablenotetext{b}{Events detected from \emph{ACE} in-situ measurements.}
\end{center}
\label{table:table2}
\end{table}

With 22,719 SFR events in the database, a thorough and careful categorization becomes essential to delineate various effects. Table \ref{table:table2} presents the detailed categorization which forms the basis for the subsequent analysis. Considering that one unique aspect of the \emph{Ulysses} orbit is the high latitude it reached (up to $\pm$ 80.2$^{\circ}$) and the role the latitude plays in the main parameters of flux ropes, we first categorize all flux ropes into two groups based on the latitudes with a separation point at 30$^{\circ}$. As listed in the Table \ref{table:table2} (also separated by grey areas in Figure \ref{fig:overview}), the time periods of these groups are comparable in length. Basically, the time periods when \emph{Ulysses} was at high latitudes, as denoted by grey areas in Figure \ref{fig:overview}, always correspond to either solar maximum or minimum. As for low latitudes, except for the several months traveling from the south pole to the north, most times are transitioning periods between these two types of extrema. The range of radial distances of the two groups is all from 1.34 to 5.41 au, which is also the distance range of the \emph{Ulysses} spacecraft. The total number of flux ropes detected at higher latitudes (greater than or equal to 30$^{\circ}$) under Wal\'en slope threshold 0.5 is 13,124, and it is 9,595 at latitudes less than 30$^{\circ}$, over both the north and south hemispheres.

The second category is based on different radial distances, i.e., $\sim$ 1 au and 3.5+ au near the ecliptic. The database at $\sim$ 1 au is obtained via the \emph{ACE} spacecraft measurements in order to achieve better statistics and to enable an examination on the radial dependence of flux rope properties. As for distances greater than 3.5 au, two specific time periods of \emph{Ulysses} mission are selected. Both of them satisfy the requirements of low latitude and far distances. The corresponding time periods at \emph{ACE} are also selected to facilitate a one-to-one comparison under otherwise similar set of conditions. The numbers of SFRs in these two groups at different radial distances are 17,620 and 10,124, respectively. Note that this comparison is implemented using different but consistent search algorithm scenario for the two groups, which will be discussed in details in Section \ref{sec:rad}.

SFRs in the third category are all detected during three fast latitude scans of \emph{Ulysses}, as enclosed by blue and dark orange dashed lines in Figure \ref{fig:ulycount}. During these periods, the ranges of latitudes and radial distances are nearly identical. Such consistency enables us to investigate the features of SFRs at solar maximum and minimum qualitatively. Since the \emph{Ulysses} mission encompasses one solar maximum period only but two solar minimum periods, the total number of events for the former is 1,238, close to the half of 2,912, the sum of the SFRs counts during the two solar minimum periods. In what follows, we will present the detailed statistical studies on each category of events, as listed in Table \ref{table:table2}.

\section{Latitudinal Effects on Small-scale Flux Ropes}\label{sec:lat}
\begin{figure}
\begin{centering}
\includegraphics[scale=0.76]{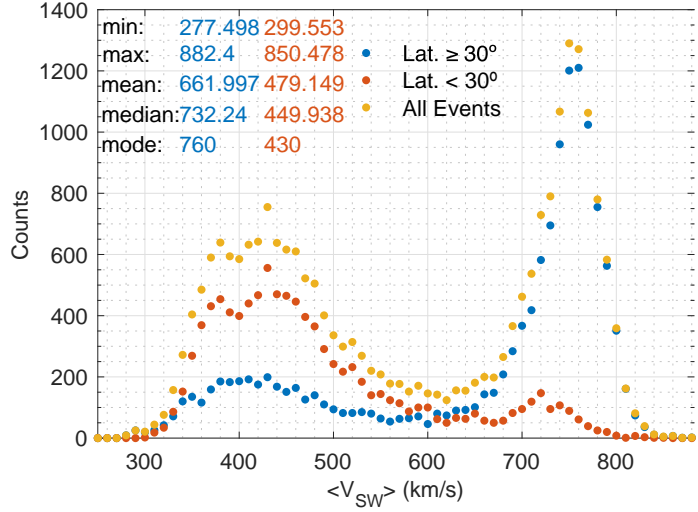}
\caption{Distribution of average solar wind speed for all small-scale flux ropes in Category I. The bin size is 10 km/s. The blue dots denote flux ropes detected at higher latitudes, whereas the dark orange dots are those at lower latitudes. The entire event set is marked by the golden dots. Various statistical quantities are also denoted.}\label{fig:vsw}
\end{centering}
\end{figure}

As indicated in Figure \ref{fig:overview}, the distribution of solar wind speed has a clear latitudinal dependence, especially during solar minima. Correspondingly, the plasma parameters, such as the proton temperature and proton number density, etc, vary accordingly. We first examine the latitudinal effects based on the Category I classification. Figure \ref{fig:vsw} presents the distributions of solar wind speed for the high and low latitude groups (blue and dark orange dots, respectively) and the entire event set (golden dots). Although the maximum values of two groups are close (882 km/s and 850 km/s), it is clear that high speed wind occurs at higher latitudes more often which leads to the larger mean (661 km/s) and median (732 km/s) for the group of events in higher latitudes. On the other hand, the group in lower latitudes tends to have relatively lower solar wind speed.

\begin{figure}
\begin{centering}
\includegraphics[scale=0.66]{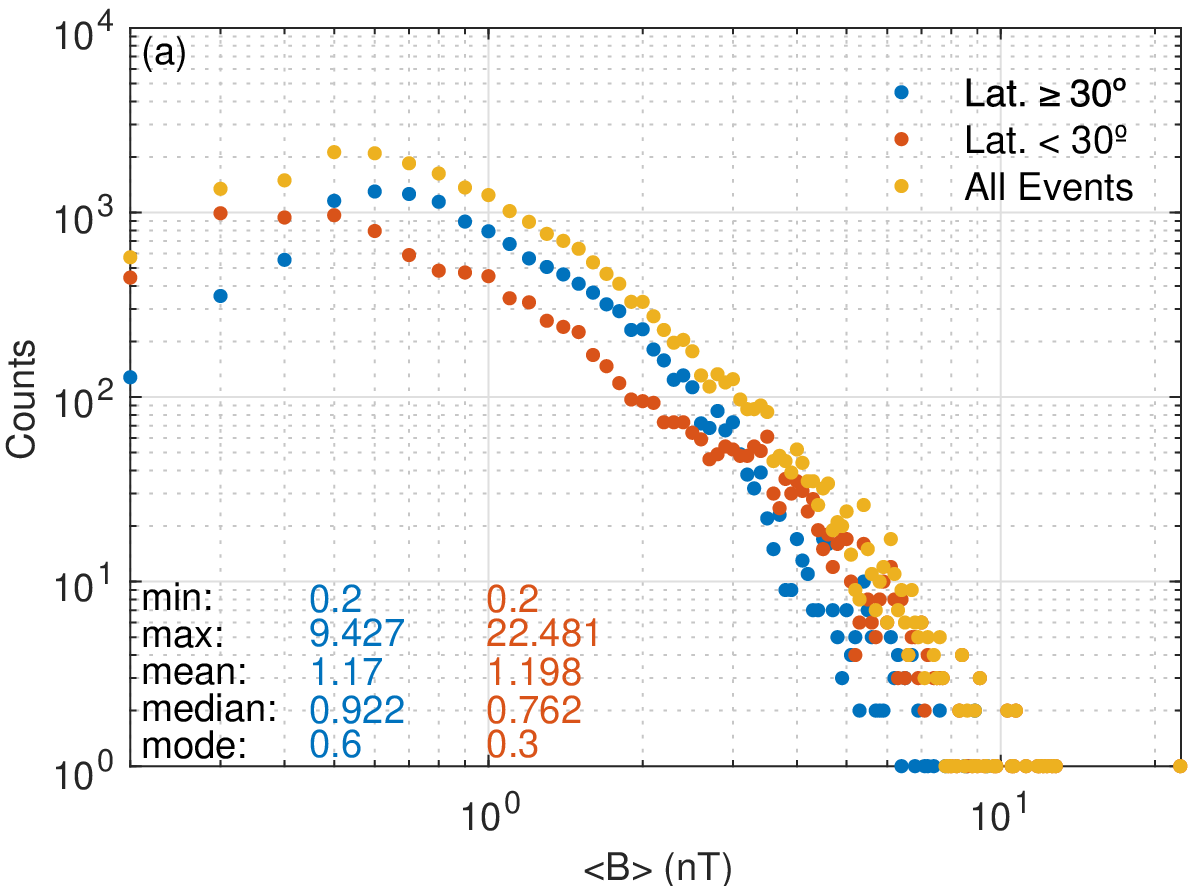}
\includegraphics[scale=0.66]{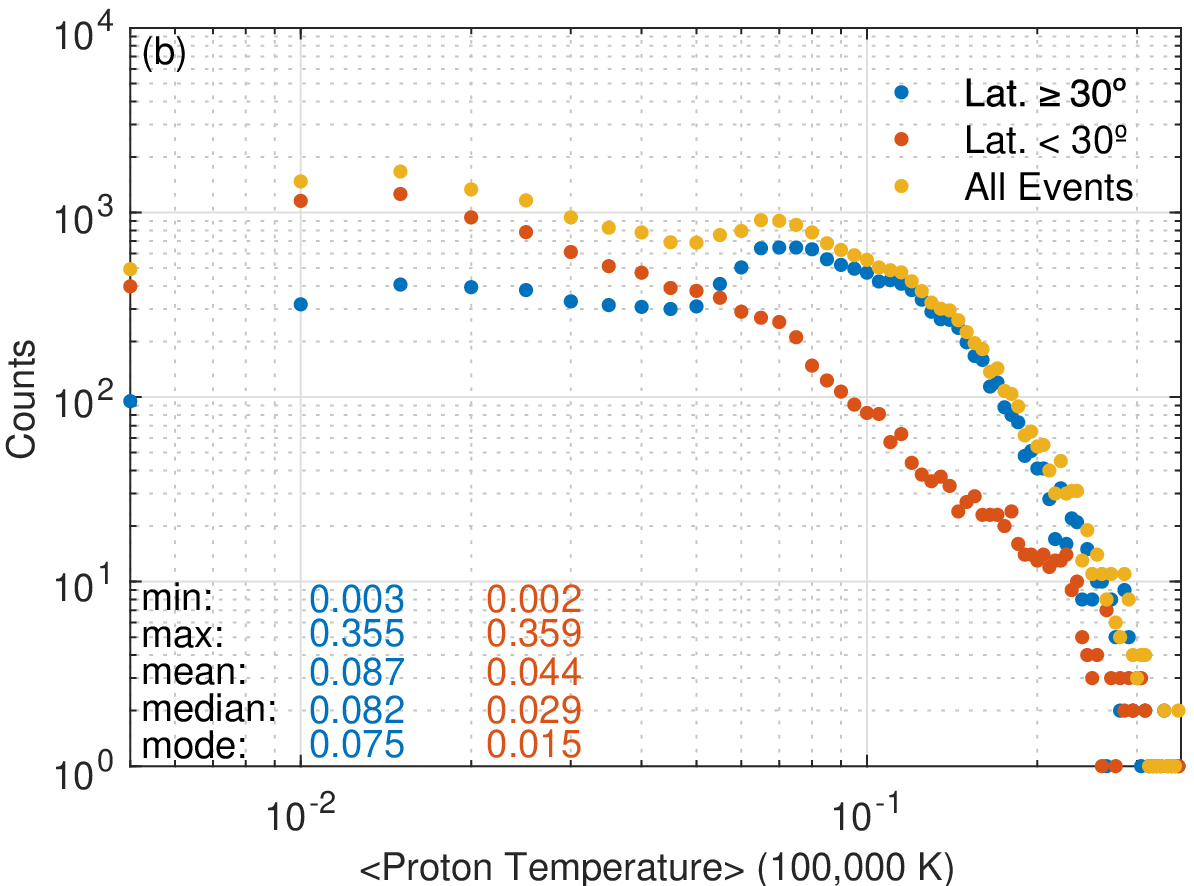}
\includegraphics[scale=0.66]{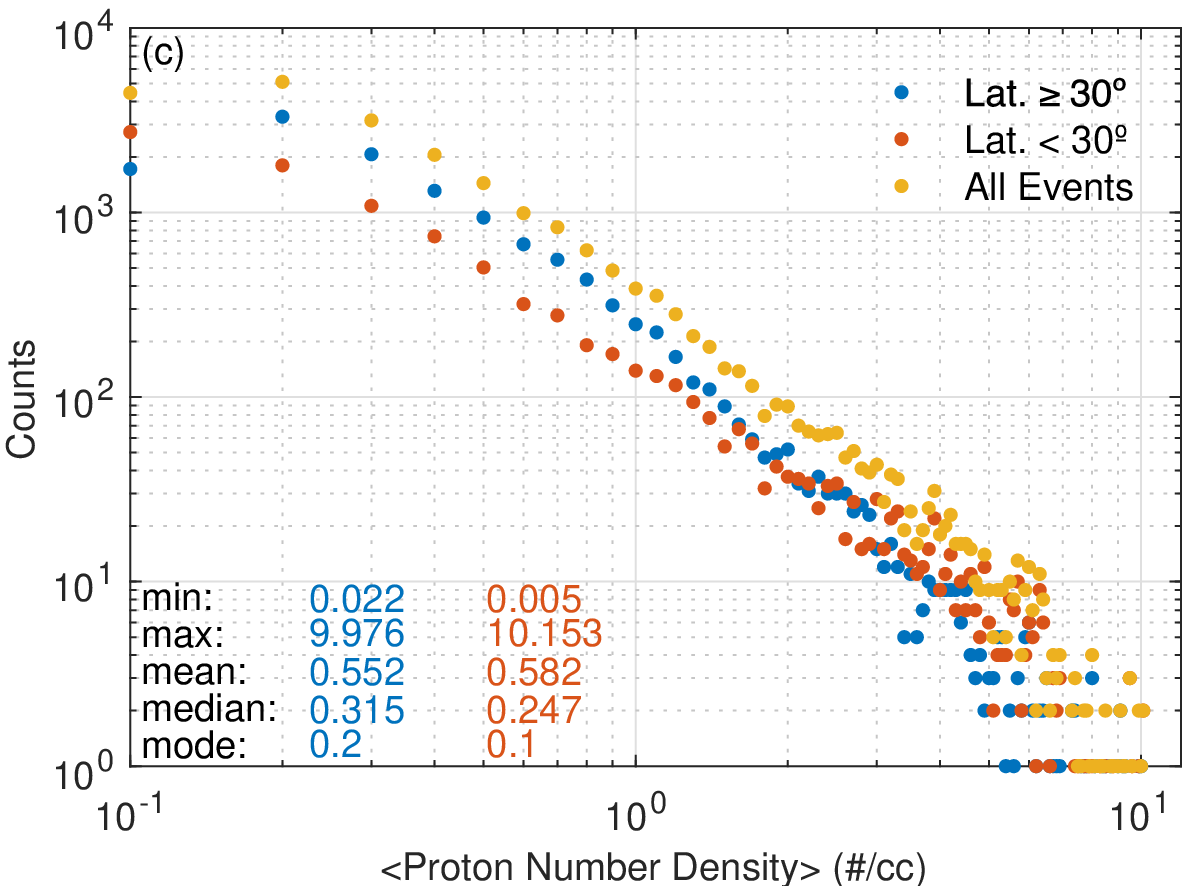}
\includegraphics[scale=0.66]{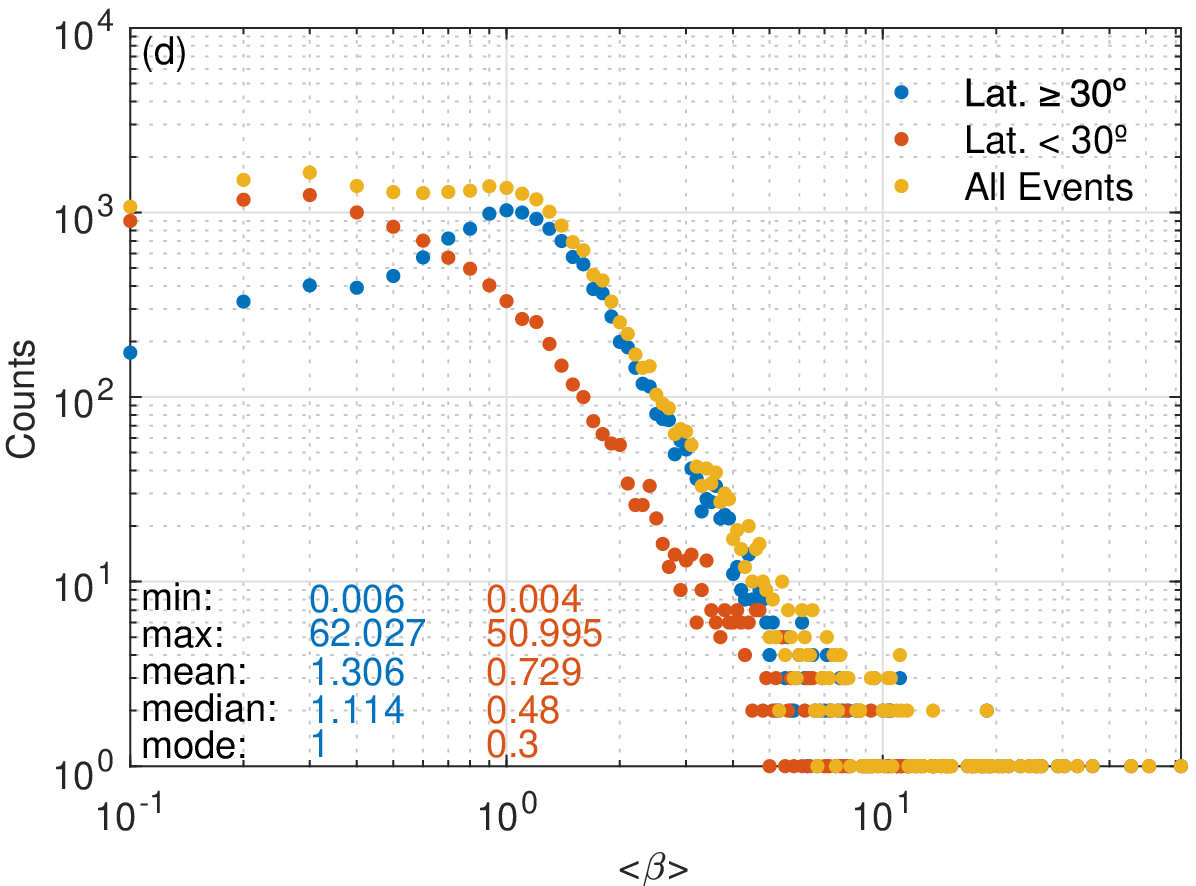}
\caption{Distributions of magnetic and plasma parameters of small-scale flux ropes for Category I: (a) The average magnetic field magnitude; bin size is 0.1 nT. (b) The proton temperature $T_p$ with 0.005 $\times$ $10^6$ K as bin size. (c) The proton number density with 0.1 $\#/{cm}^3$ as bin size. (d) The average proton beta with 0.1 as bin size. The format is the same as Figure \ref{fig:vsw}.}\label{fig:ulyparameter}
\end{centering}
\end{figure}

Recall that in the summary plot of the temporal variations, Figure \ref{fig:overview}, the extreme values of almost all parameters are within the periods when the \emph{Ulysses} passed through the ecliptic plane at the closest radial distances. In reality, each is a short period (2 $\sim$ 3 months) for \emph{Ulysses} traveling from $\leqslant$ -30$^{\circ}$ to $\geqslant$ 30$^{\circ}$ latitude at a radial distance $\sim$ 1.3 au. The total number of flux ropes during these time periods is 952, less than 10\% of the event count in the low latitude group. Consequently, these records will not affect too much on the overarching distributions except for the scatter mostly seen toward the tail (maximum values) of each distribution.

Figure \ref{fig:ulyparameter} shows the distributions of plasma and magnetic field parameters for each flux rope interval. Figure \ref{fig:ulyparameter}a indicates that the magnetic field magnitude increases sometimes with latitudinal decrease, but such possible latitudinal effect on magnetic field is not significant compared with the effect corresponding to the changing radial distances (see Section \ref{sec:rad}). The extreme values in the corresponding distribution are related to the existence of the smallest and largest radial distances in that group. Although the minima of both groups are equal, the maximum magnitude of the low latitude group (22.5 nT) is at least twice as that of the high latitude group (9.4 nT). Figures \ref{fig:ulyparameter}b-c also verify that this explanation can be applied to the distributions of the proton temperature $T_p$ and proton number density $N_p$. Both the maximum and minimum values are included in the group of low latitudes. However, the maxima of two groups are actually close, which means that radial distance may not have that much influence as it does on magnetic field. Furthermore, the difference between two distributions of $T_p$ (blue and dark orange dots) is quite significant, which is owing to the fact that the flux ropes at higher latitudes are inclined to have high proton temperature due to relatively high speed. Figure \ref{fig:ulyparameter}d is the distribution of plasma $\beta$. By definition, plasma $\beta$ combines all three aforementioned parameters. It shows that there are more high $\beta$ values appearing at higher latitudes, most likely caused by the enhanced $T_p$ shown in Figure \ref{fig:ulyparameter}b.

\begin{figure}
\begin{centering}
\includegraphics[scale=0.66]{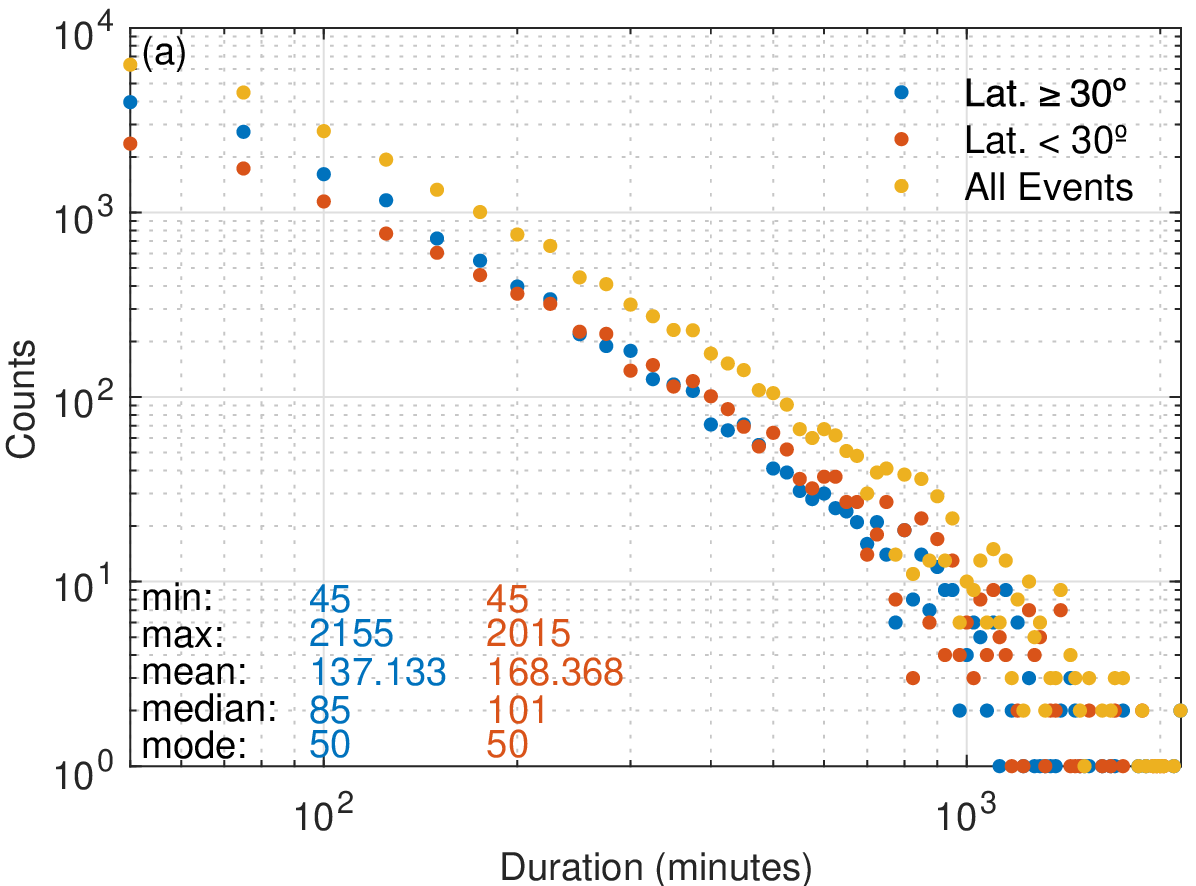}
\includegraphics[scale=0.66]{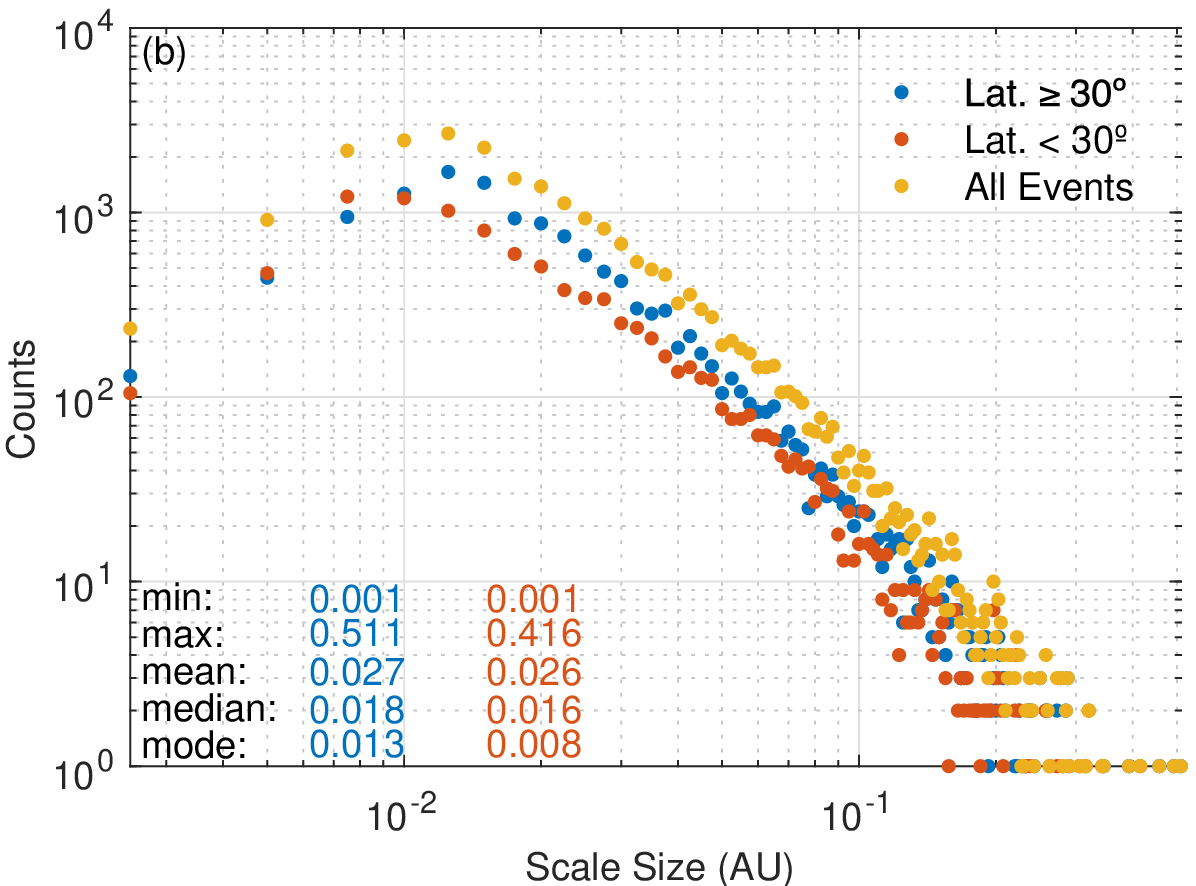}
\includegraphics[scale=0.66]{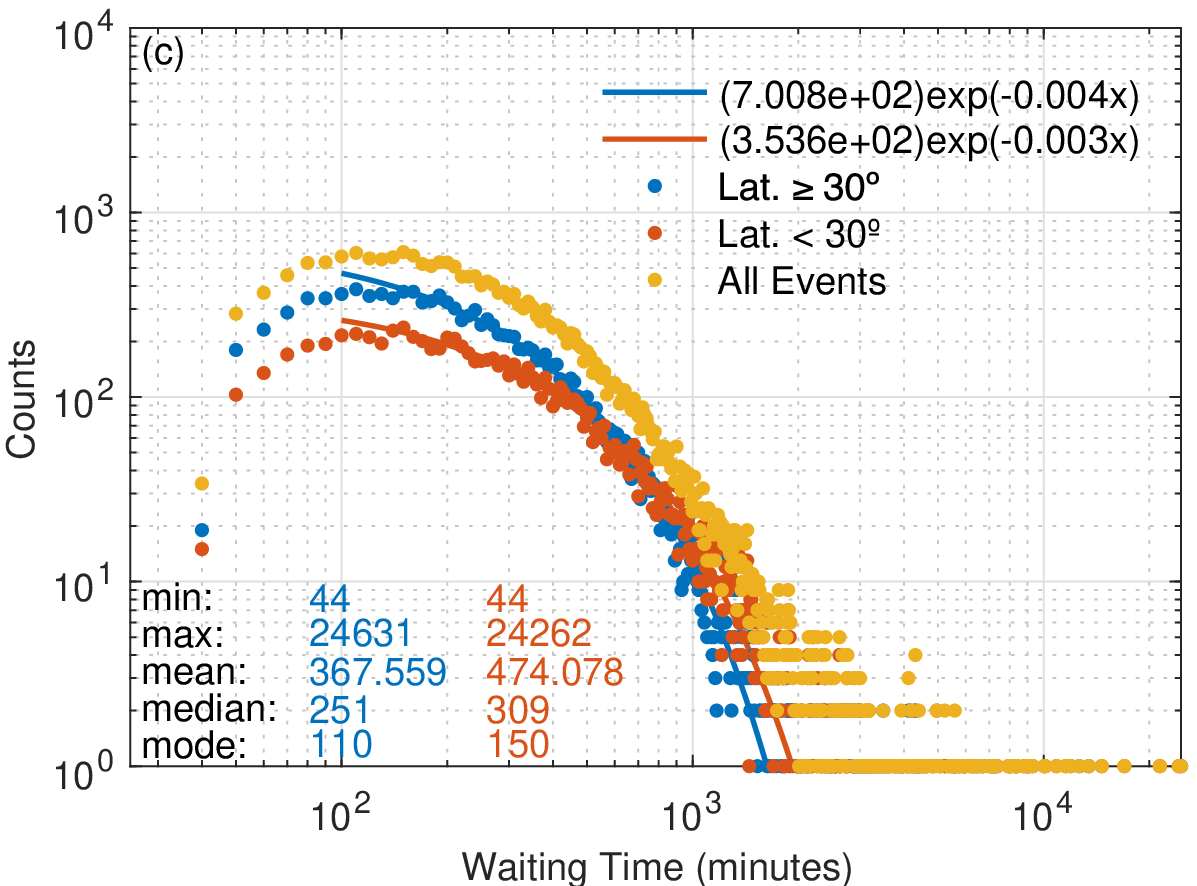}
\includegraphics[scale=0.66]{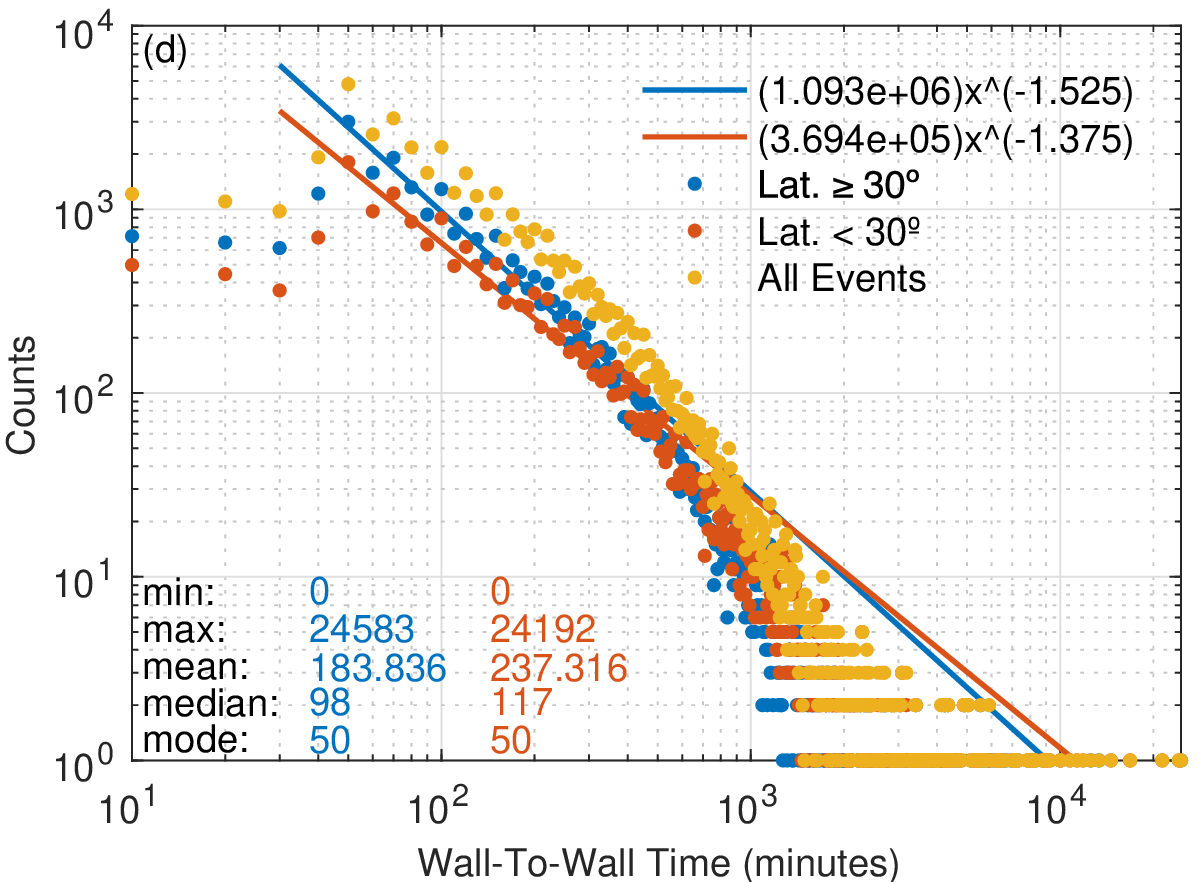}
\caption{Distributions of small-scale flux rope properties for Category I: (a) Duration with 30 minutes as bin size. (b) Scale size with 0.0025 au as bin size. (c) The waiting time with two exponential fitting curves as denoted by the solid curves, 10 minutes as bin size. (d) The wall-to-wall time distributions with power-law fitting curves for the portion of each group as denoted, the same bin size as (c). The format is the same as Figure \ref{fig:vsw}.}\label{fig:ulyparameter1}
\end{centering}
\end{figure}

The distributions related to flux rope duration are presented in Figure \ref{fig:ulyparameter1}. The duration is the time span between the start and end times of each flux rope interval. Figure \ref{fig:ulyparameter1}a indicates that two groups of flux ropes have similar power-law like distributions in duration, but the statistical quantities are slightly different, in terms of the mean (137 and 168 minutes, respectively) and the median (85 and 101 minutes, respectively). The scale size is calculated along the projection of the spacecraft path onto the flux rope cross section which lies on the plane perpendicular to the flux rope axis. Again, the distributions of two groups as presented in Figure \ref{fig:ulyparameter1}b are close with nearly identical mean values, 0.027 and 0.026 au, respectively. Both parameters exhibit power-law distributions for the two groups, but they seem to possess different power-law slopes (indices).

Figure \ref{fig:ulyparameter1}c is obtained from the waiting time analysis which can be utilized to determine whether discrete events occur independently \citep{Pearce1993}. In our study, the waiting time is defined by the elapsed time between two starting times of the adjacent flux rope intervals. Both groups obey the exponential function fittings as shown, but with slightly different fitting exponents. Figure \ref{fig:ulyparameter1}d is the distribution of wall-to-wall time \citep{Zheng2018}. By ``wall'' we mean the current sheet with zero thickness that exists at the boundary of each flux rope. Therefore, the wall-to-wall time is equivalent to the separation between (waiting time of) these current sheets. The power law fittings are applied to both groups and plotted as solid lines in the corresponding colors for the portions with the most significant number of counts corresponding to a range of waiting times between a few tens to a few hundreds minutes. We omit the portions beyond certain break points (at $\sim$ 300 minutes) for both groups, beyond which the counts are relatively low and the distributions seem to steepen, not to clutter the plot.

\begin{figure}
\begin{centering}
\includegraphics[scale=0.66]{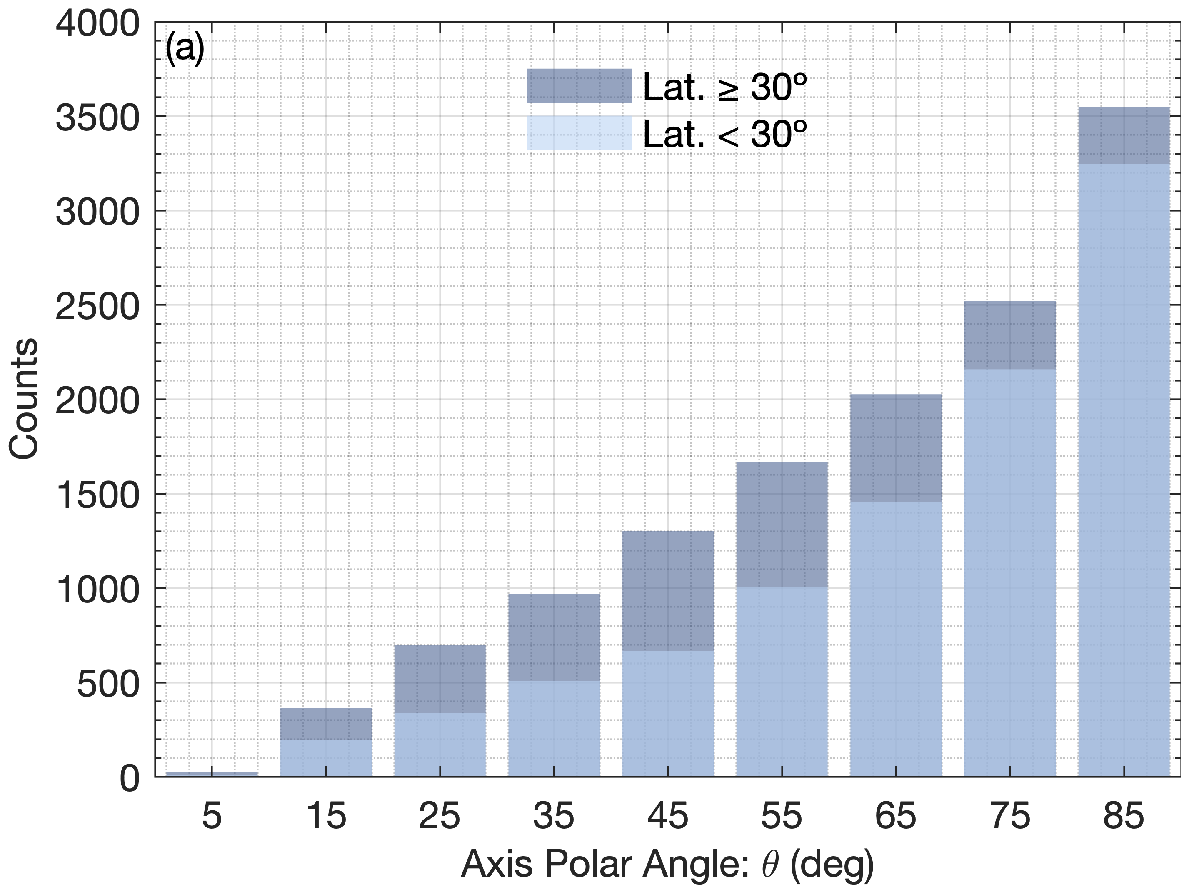}
\includegraphics[scale=0.66]{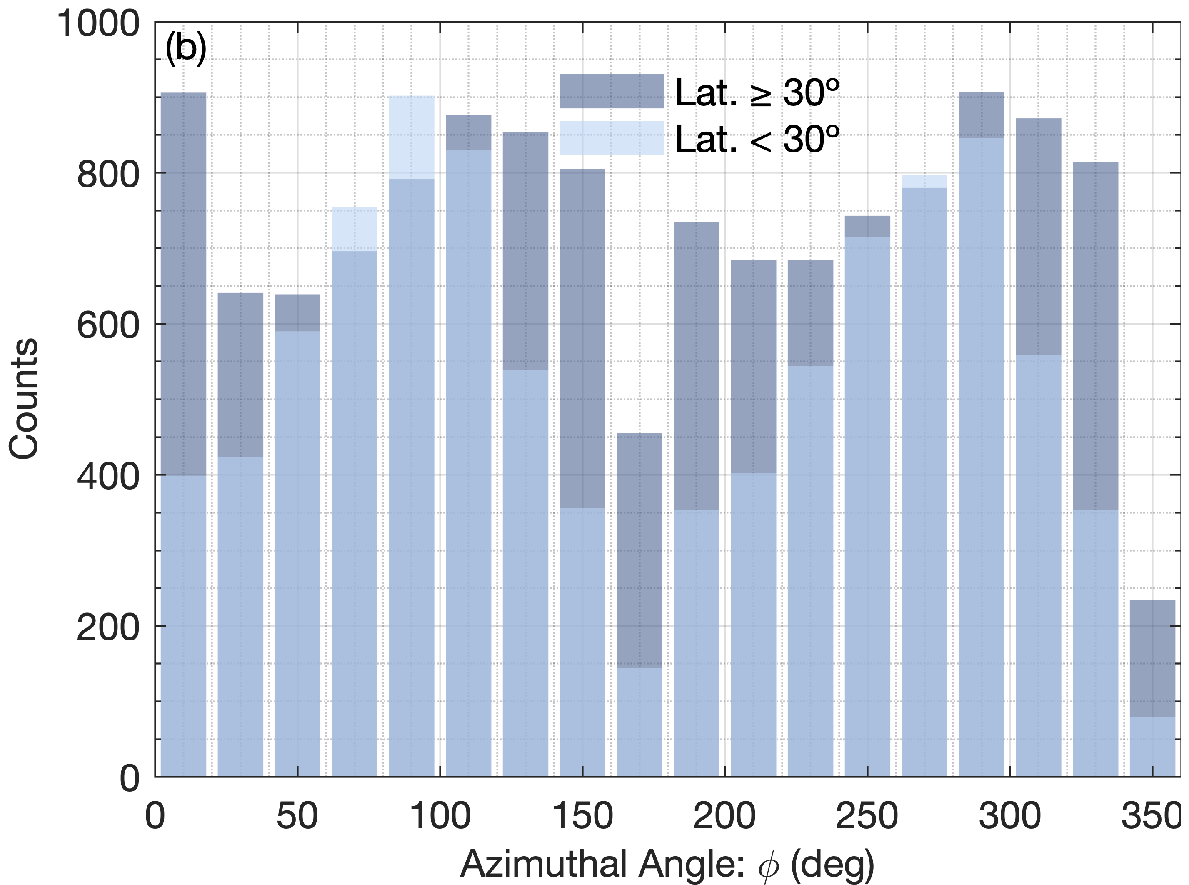}
\caption{Distribution of the axis orientation of flux ropes at different latitudes: (a) The axis polar angle $\theta$ with 10$^{\circ}$ as bin size. (b) The azimuthal angle $\phi$ with 20$^{\circ}$ bin size. The dark blue bars are for high latitudes flux ropes while the light blue ones represent those at lower latitudes. }\label{fig:ulyparameter2}
\end{centering}
\end{figure}

In addition, there is one more set of parameters of flux rope characteristics, i.e., the $z$-axis orientation. Figure \ref{fig:ulyparameter2} presents the distributions of their angular directions in $(\theta, \phi)$ angles. Figure \ref{fig:ulyparameter2}a is the distribution of $\theta$, the angle between the flux rope $z$-axis (given in the $RTN$ coordinate system) and the local $N$ direction. Flux ropes at both high and low latitudes have peaks near 85$^{\circ}$, which is evidence that most of SFRs tend to lie on the local $RT$ plane (near ecliptic for low latitudes). The azimuthal angle, $\phi$, measures the angle between the projection of $z$-axis onto the $RT$ plane and the $R$ direction. Figure \ref{fig:ulyparameter2}b suggests that flux ropes at low latitudes have two peaks of $\phi$ ($\sim$ 90$^{\circ}$ and $\sim$ 290$^{\circ}$), whereas the peaks for those at higher latitudes are less pronounced. Although two groups have dissimilar distributions, most of the flux ropes are still aligned with the Parker spiral for the low latitude group. Notice that the angle of the Parker spiral is no longer a simple value such as the 1 au result near the ecliptic reported in \cite{Hu2018}, it is a function of radial distance, solar wind speed, and heliographic latitude instead when considering to the unique orbit of the \emph{Ulysses} spacecraft, especially with the rapid latitudinal variation \citep{Balogh2001}.

\begin{figure}
\begin{centering}
\includegraphics[scale=0.66]{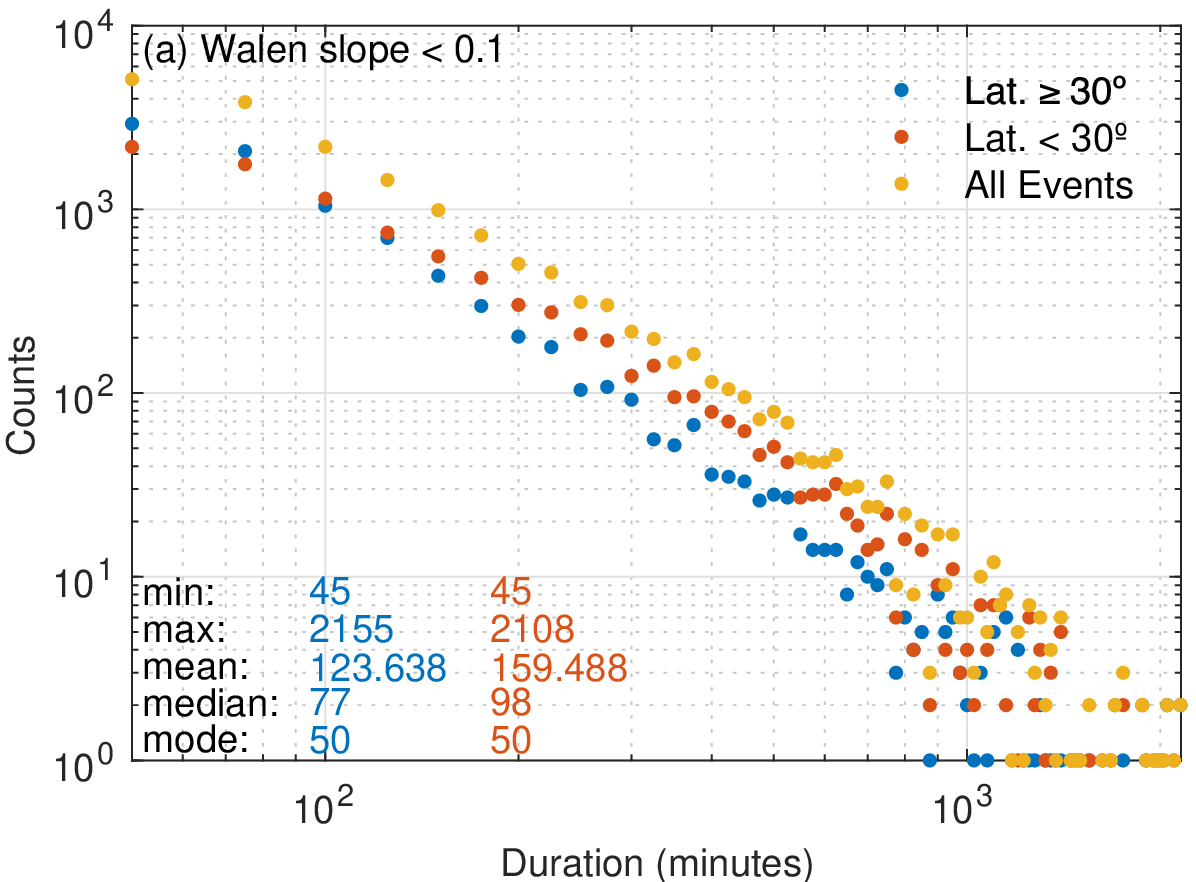}
\includegraphics[scale=0.66]{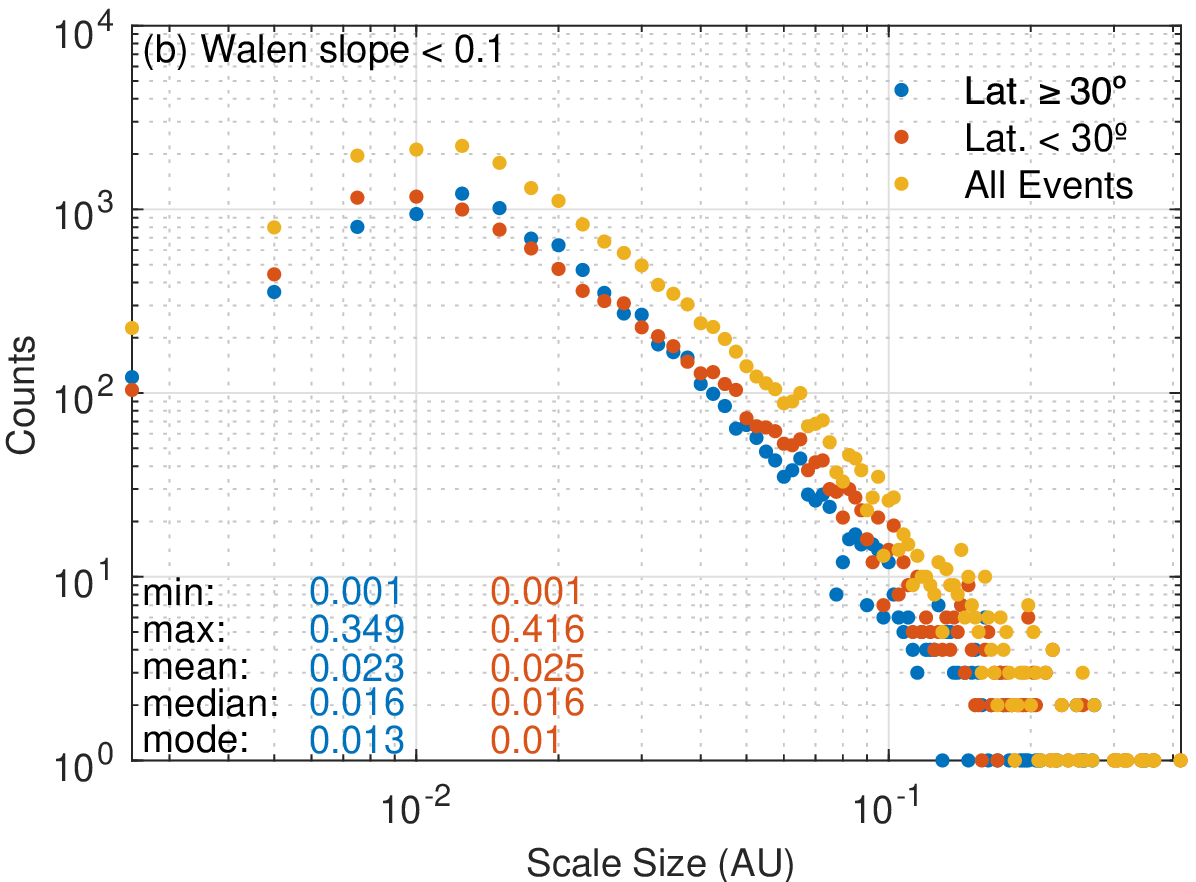}
\includegraphics[scale=0.66]{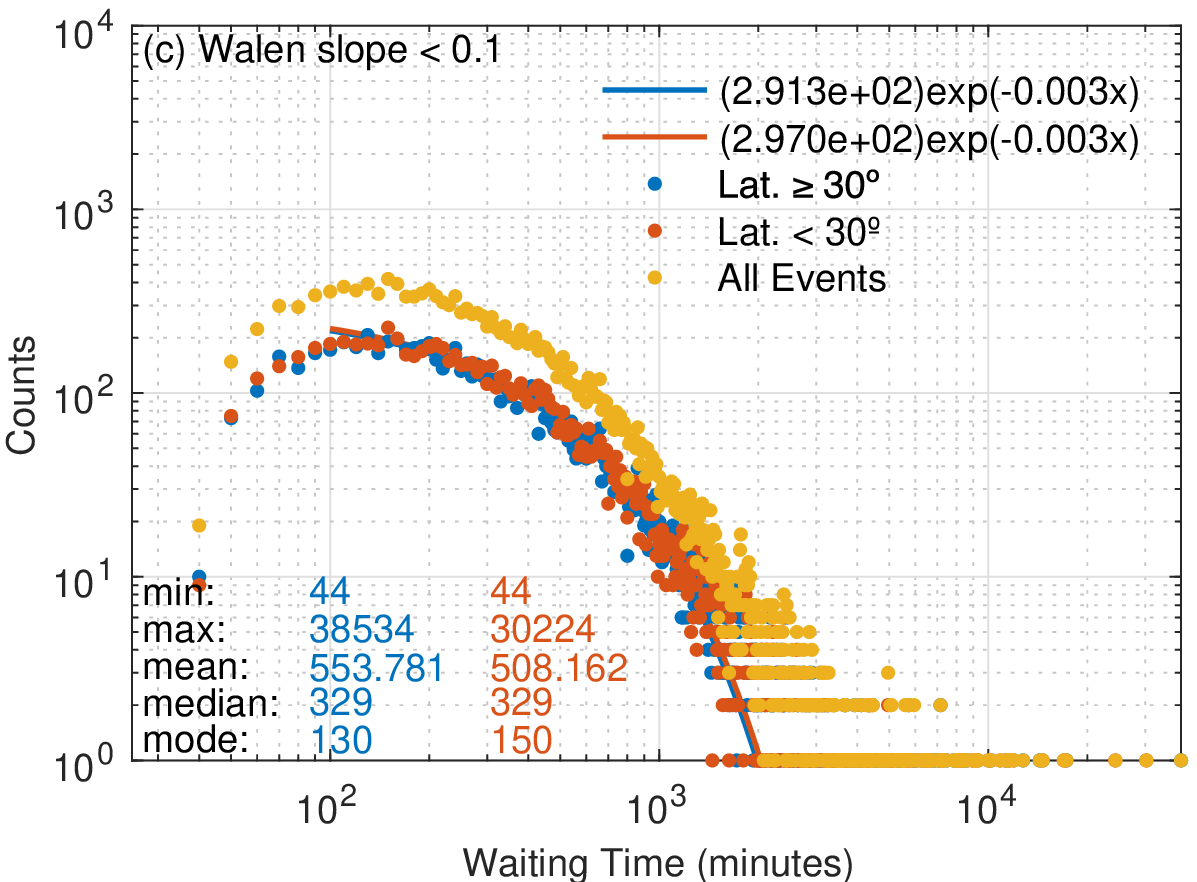}
\includegraphics[scale=0.66]{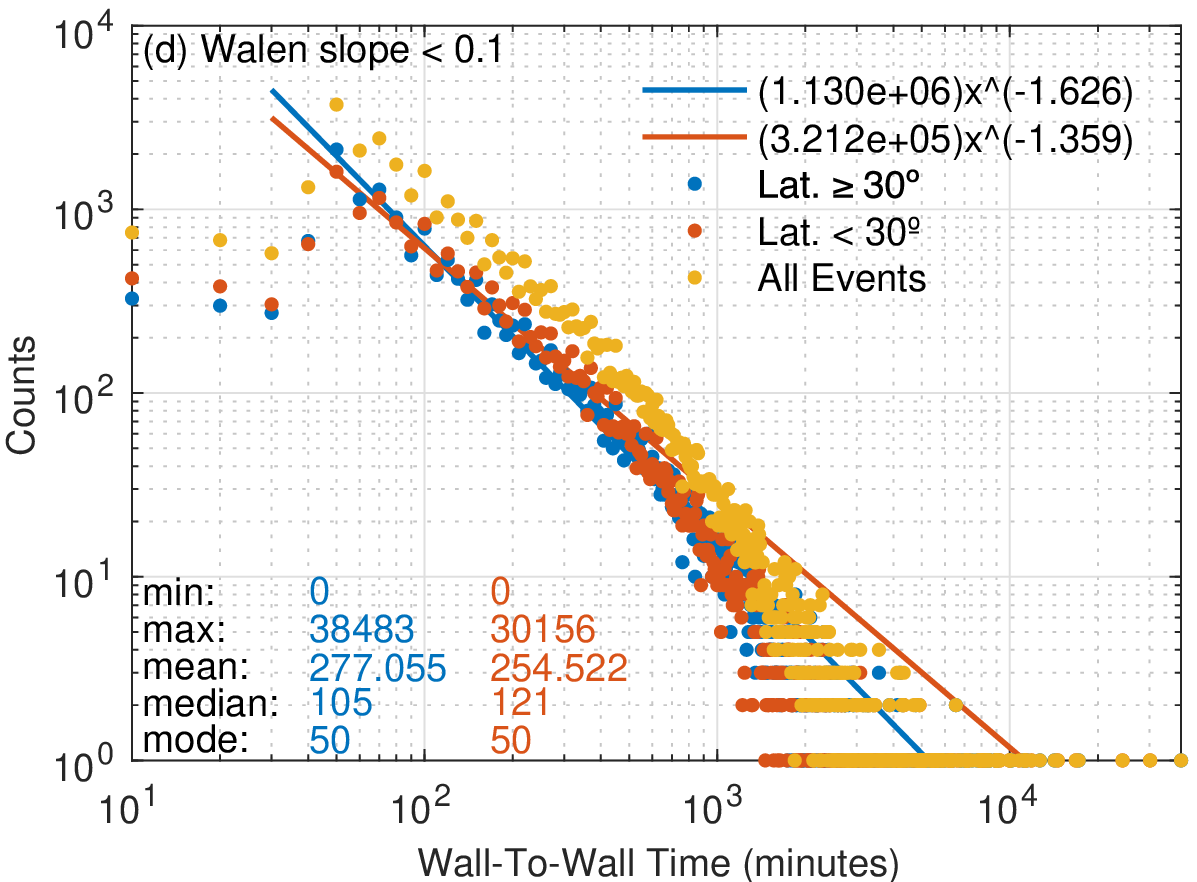}
\caption{Distributions of small-scale flux rope properties for Category I, with the Wal\'en slope threshold, 0.1. The format is the same as Figure \ref{fig:ulyparameter1}.}\label{fig:ulyparameterr1}
\end{centering}
\end{figure}

As mentioned in Section \ref{sec:overview}, due to the existing Alfv\'enic structures or waves, we can lower the Wal\'en slope threshold from 0.5 to 0.1, in order to exclude more strictly these possible Alfv\'en-wave-like structures. After applying this Wal\'en slope threshold, 0.1, the low latitude group now has 8,950 events, whereas the high latitude group has 8,711 events. Again, these numbers suggest that flux ropes with high Wal\'en slope numbers occur more often at high latitudes, and the latitudinal effect on SFRs has been removed to a great extent since the differences between different latitudes become minimal as presented in Figure \ref{fig:ulyparameterr1}, especially for the distributions of waiting time. However, the distributions for the other properties remain similar to the results shown in Figure \ref{fig:ulyparameter1} in that both groups still exhibit power-law distribution with noticeably different power-law slopes. This indicates that for more strictly identified ``pristine'' flux ropes the statistical distributions of the main properties as represented by Figures \ref{fig:ulyparameter1} and \ref{fig:ulyparameterr1} do not change significantly, although the number of events is reduced significantly, especially for the high latitude group. This leads to the need of examining the dependence on radial distances, which effect may still be embedded in the present Category I.

\section{Effects of Radial Distances on Small-scale Flux Ropes}\label{sec:rad}
In addition to latitudinal effects, the radial distance is also a vital factor which would also affect the properties of flux ropes, such as the magnetic field, plasma and other derived parameters. In order to evaluate how important the radial dependence is, the events in the database of flux ropes around 1 au are employed to facilitate a direct comparison with the corresponding \emph{Ulysses} events at radial distances greater than 3.5 au. Before we move on to detailed comparisons, it is necessary to discuss the additional approaches we take to address the issues caused by different data quality and the consequences in constructing the two event subsets belonging to Category II.

\subsection{Comparison of Two Scenarios of Detection Results}
\cite{Zheng2018} published the database and analysis of small-scale flux ropes via the \emph{Wind} spacecraft measurements with 1 minute cadence. Due to very different resolutions of the magnetic field and plasma parameters, the detection for the \emph{Ulysses} datasets cannot be repeated with the uniform 1 minute resolution data through simple data interpolation. In order to have a comparison at different radial distances under the same conditions, here we adopt an alternative approach for calculating the transverse pressure $P_t$, which bridges the gap between the two databases. As introduced in Section \ref{sec:gs}, the algorithm of calculating $P_t$ includes the thermal pressure $N_pkT_p$ and the axial magnetic pressure ${B_z}^2/{2\mu_0}$. Alternatively, considering that the 1 minute cadence magnetic field data are available from\emph{Ulysses} while the plasma parameters and solar wind speed are 4 to 8 minute averages, we neglect the thermal pressure in calculating the transverse pressure which requires the use of magnetic field data only.

\begin{figure}
\begin{centering}
\includegraphics[scale=0.66]{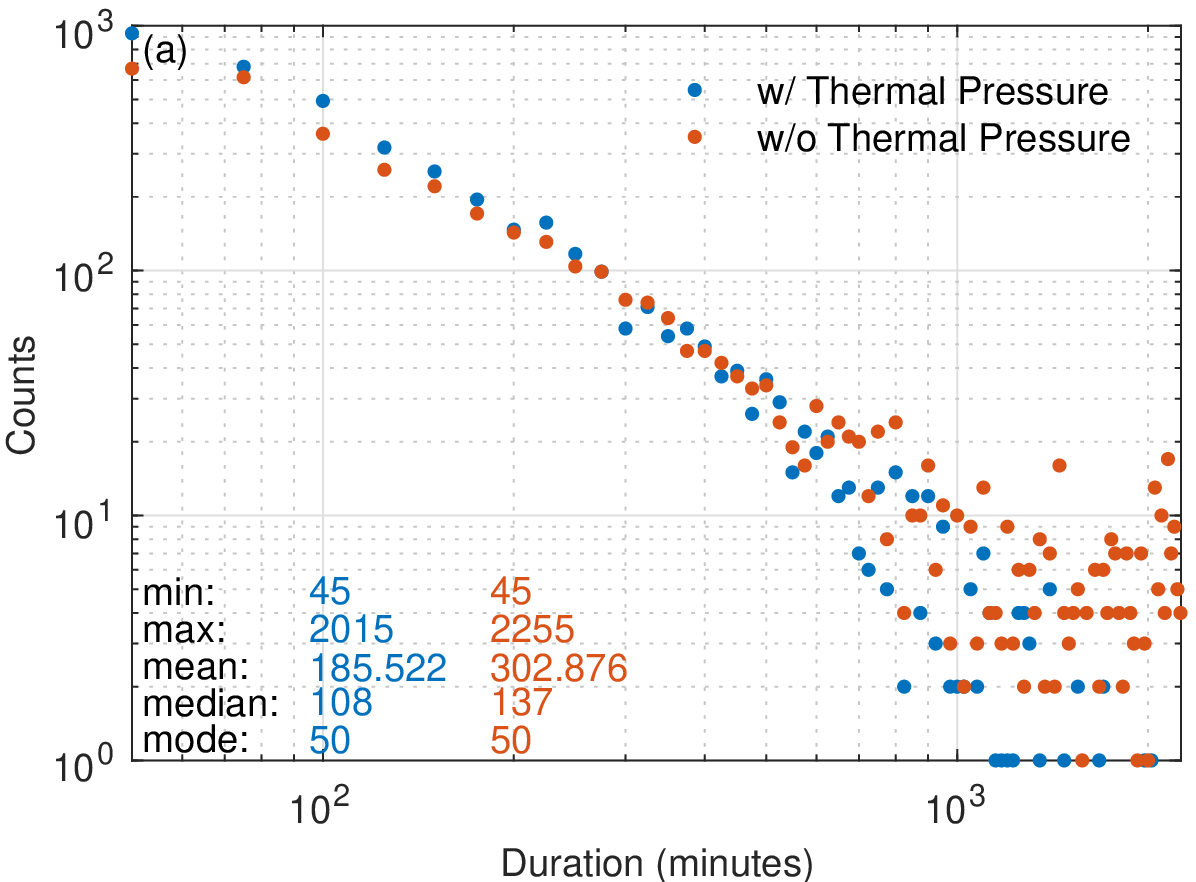}
\includegraphics[scale=0.66]{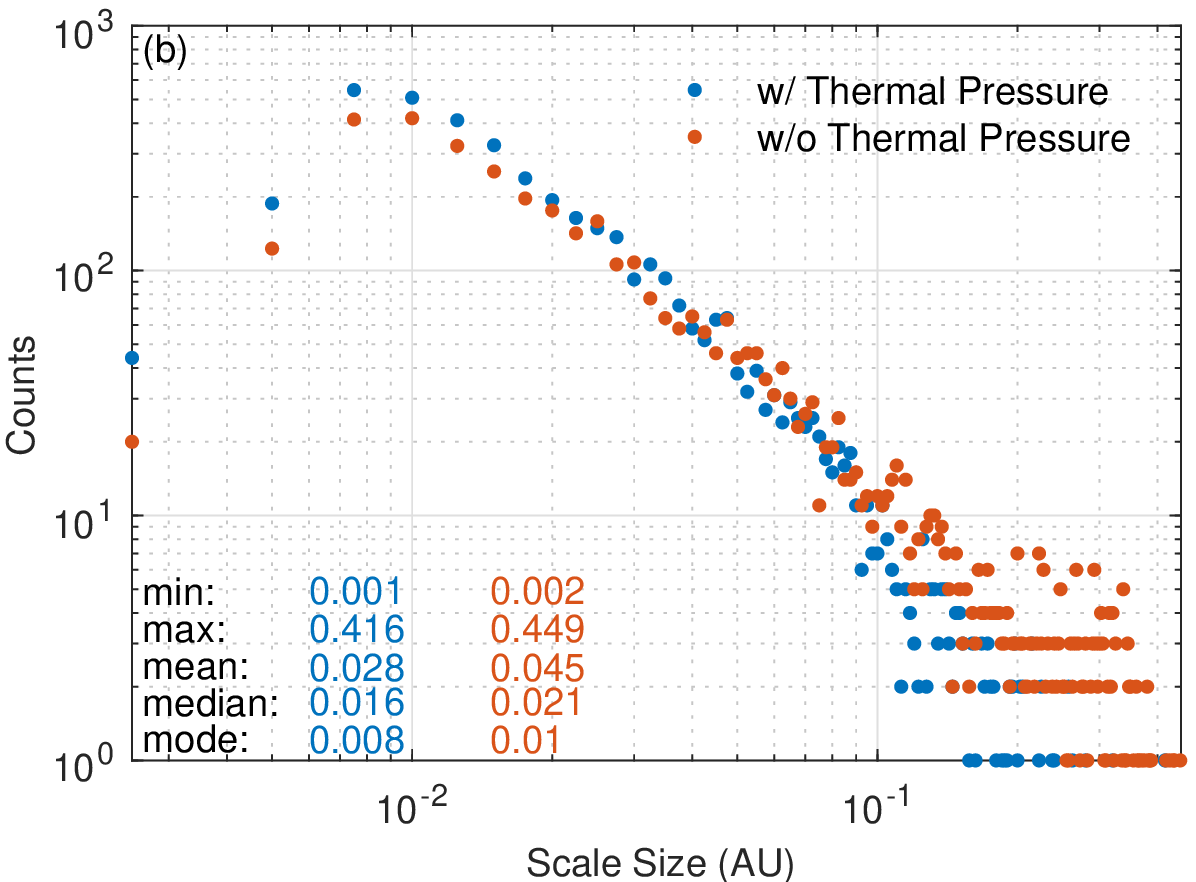}
\includegraphics[scale=0.66]{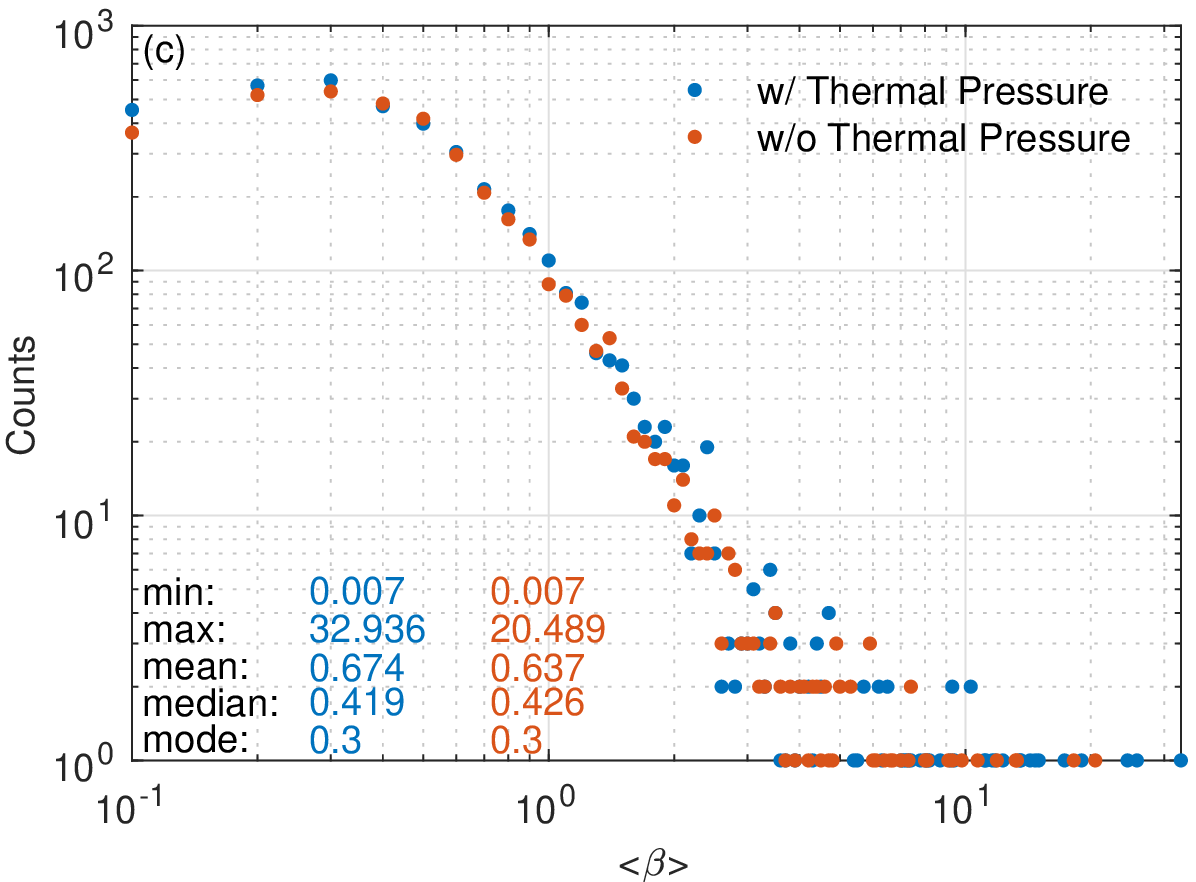}
\includegraphics[scale=0.66]{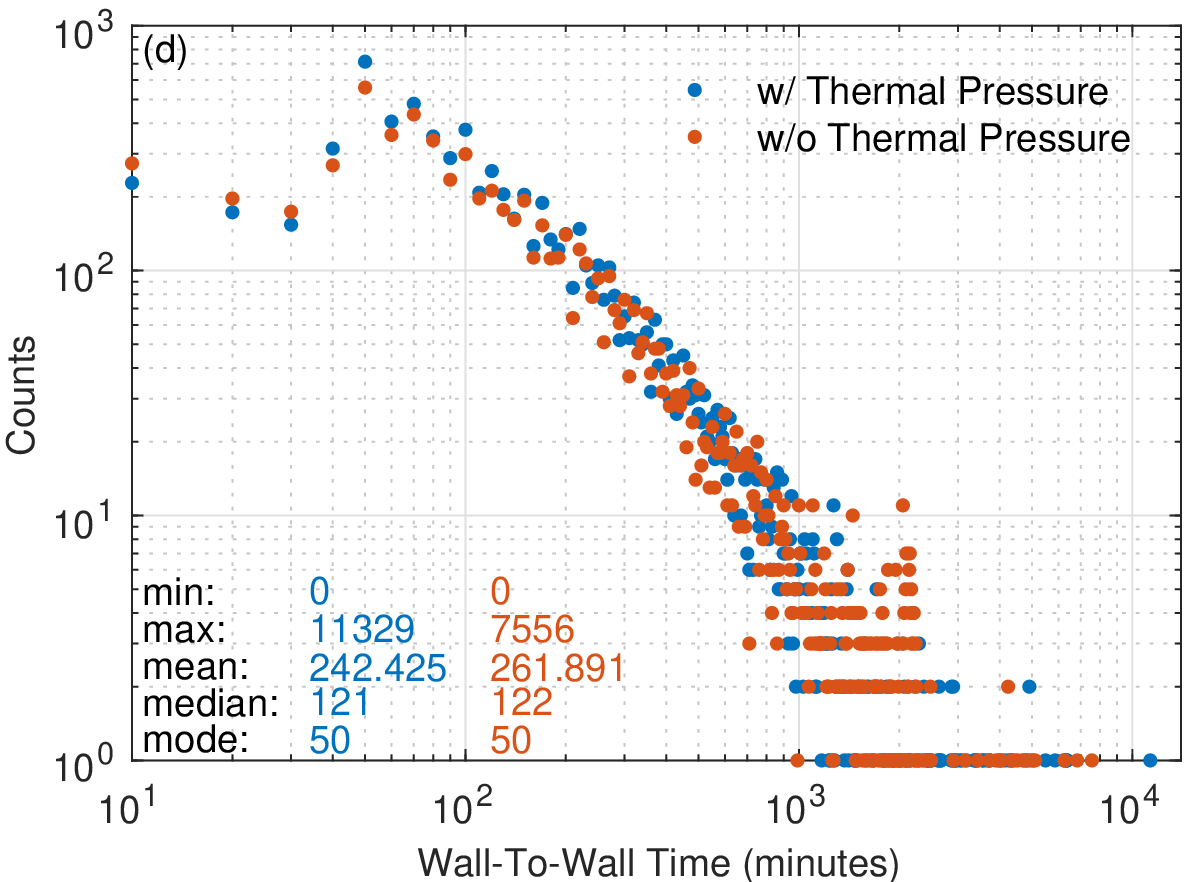}
\caption{Distributions of small-scale flux ropes via \emph{Ulysses} measurements under two algorithm scenarios (with and without thermal pressure): (a) Duration with 25 minutes as bin size. (b) Scale size with 0.0025 au as bin size. (c) The average plasma $\beta$ with 0.1 as bin size. (d) The wall to wall time with 10 minutes as bin size. The blue dots represent searching results with thermal pressure while the dark orange dots represent results without it.}\label{fig:ulypressure}
\end{centering}
\end{figure}

With the same set of criteria shown in Table \ref{table:table1} (note that the minimum duration allowed is 45 minutes), Figure \ref{fig:ulypressure} presents the comparison of detection results based on these two scenarios, one with thermal pressure and one without in the calculations of $P_t$. With the thermal pressure included, we detected 4,090 SFRs, whereas number without thermal pressure is 3,798. As seen from Figure \ref{fig:ulypressure}, the difference in the statistical distributions between these two scenarios is negligible, especially by excluding the low-count portions toward the tails. Since the main reason for us to adopt the new scenario is to compare this database with 1 au detection results, the time periods in Category II (Table \ref{table:table2}) are selected so that one-to-one correspondence can be established between the two sets of results obtained near the ecliptic but at different radial distances.

\subsection{Database of Small-scale Flux Ropes around 1 au}
Following the previous study by \cite{Zheng2018}, we extend the automated detection to \emph{ACE} spacecraft measurements covering the time period from February 1998 to the end of 2017. The searching criteria are not exactly the same as those for the \emph{Ulysses} database. Instead of starting from (45, 80) minutes, the search window range begins at (9, 16), (14, 21), and runs up to an upper limit of 2,165 minutes since 1 minute cadence data are available. The thresholds of two residues are the same as those for the \emph{Ulysses} detection, i.e., 0.12 and 0.14, respectively (see Table \ref{table:table1}). The Wal\'en slope threshold is set as 0.3. A lower limit on the magnetic field magnitude (greater than 5 nT) is applied so as to exclude small fluctuations.

\begin{figure}
\begin{centering}
\includegraphics[scale=1.0]{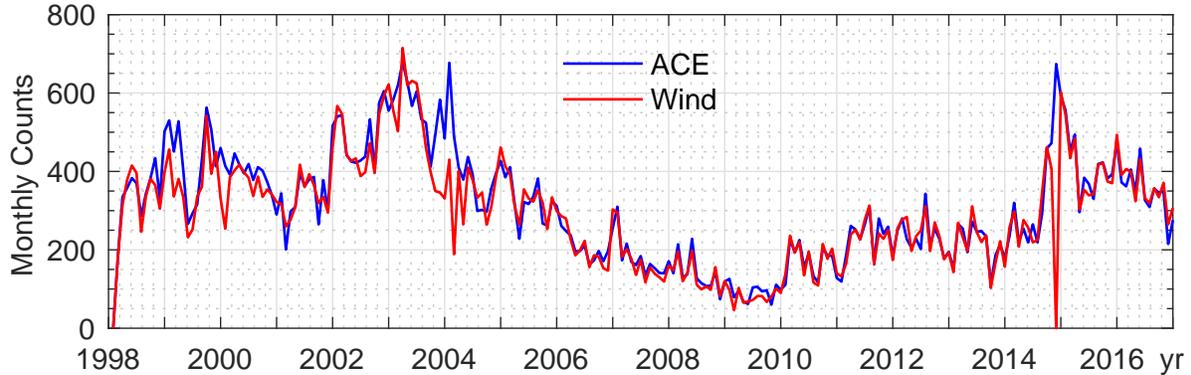}
\caption{Comparison of monthly occurrence rates of small-scale flux ropes detected via \emph{Wind} and \emph{ACE} spacecraft measurements.}\label{fig:acewindcount}
\end{centering}
\end{figure}

To verify the characteristics of SFRs at 1 au, we briefly compare and discuss the \emph{ACE} and \emph{Wind} measurements. All events lists are posted online at \url{http://www.fluxrope.info}. Following the same criteria, we first expand the \emph{Wind} database by setting duration range as 9 $\sim$ 2,165 minutes (the original is only up to 361 minutes). The comparison of SFRs between the two spacecraft are from February 1998 to the end of 2016. Alternatively, as discussed in the previous subsection regarding two scenarios of the detection algorithm, here we also choose the one without thermal pressure to compare the databases of \emph{ACE} and \emph{Wind} since there is merely slight difference between the two scenarios at $\sim$ 1 au. During the 18 years time period, we have 66,424 SFRs via \emph{Wind} and 47,249 via \emph{ACE} detection. The predominant difference between the two databases is ascribed to large data gaps of $N_p$ from \emph{ACE}. Although the scenario without thermal pressure avoids involving $N_p$ into the calculation of $P_t$, the proton number density is still needed for calculating the Alfv\'en velocity. When the Alfv\'en velocity is missing, by default none of the flux rope candidates is able to pass the Wal\'en slope test. Therefore, we perform interpolation for small data gaps and substitute a nominal value 5 $cm^{-3}$ for $N_p$ when the entire segment of data is missing. With this manual correction, the total count of \emph{ACE} events becomes 70,072 which is close to \emph{Wind} result under the same conditions as shown in Figure \ref{fig:acewindcount}. The event occurrence rate follows one and the other closely. The main noticeable difference is around 2004 when the \emph{Wind} spacecraft crossed the earth's magnetotail and in November 2014 when the entire month-long data are missing from \emph{Wind}. As expected, the basic properties of these two databases, such as the duration, scale size and wall-to-wall time distributions, etc., are nearly identical (not shown).

\subsection{Comparison of Databases between \emph{ACE} and \emph{Ulysses}}
As discussed above, the basic properties of small-scale flux ropes are identical at around 1 au between the \emph{ACE} and \emph{Wind} databases. For this reason, we then choose one database, i.e., from \emph{ACE}, to represent the detection result at near 1 au heliocentric distance and compare with detection result at far distances from the \emph{Ulysses} database. Again, considering that there is little impact from thermal pressure and more importantly, in order to extend the search window to the lower limit of 9 minutes for both spacecraft (thus only magnetic field data from \emph{Ulysses} can be used), both databases in this subsection are obtained via calculations without thermal pressure included. In order to compare these two databases in a more strict way, we select the time ranges to be from early February 1998 to mid July 1999 and from mid October 2002 to mid September 2005 for both databases as listed in Table \ref{table:table2}. The same intervals at both spacecraft are chosen because then these events can be considered to be radially aligned and the possible radial evolution may be examined. Now, the comparison is between the flux ropes at 1 au and those at radial distances greater than 3.5 au near the ecliptic only. We keep the original \emph{ACE} duration range since there probably exists diffusive effects of flux ropes from 1 au to deep space. The duration range is completely identical for the two databases, i.e., with the lower and upper limits of 9 and 2,255 minutes, respectively.

\begin{figure}
\begin{centering}
\includegraphics[scale=0.66]{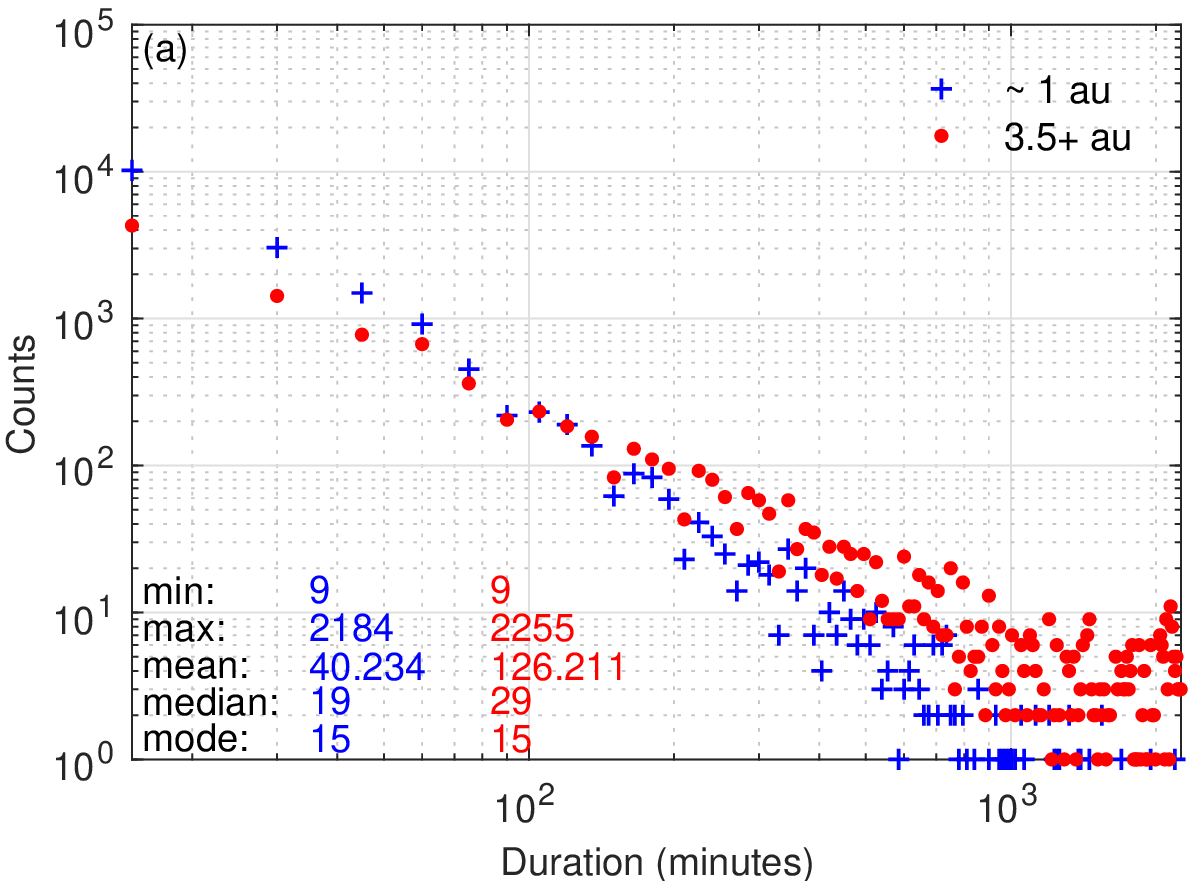}
\includegraphics[scale=0.66]{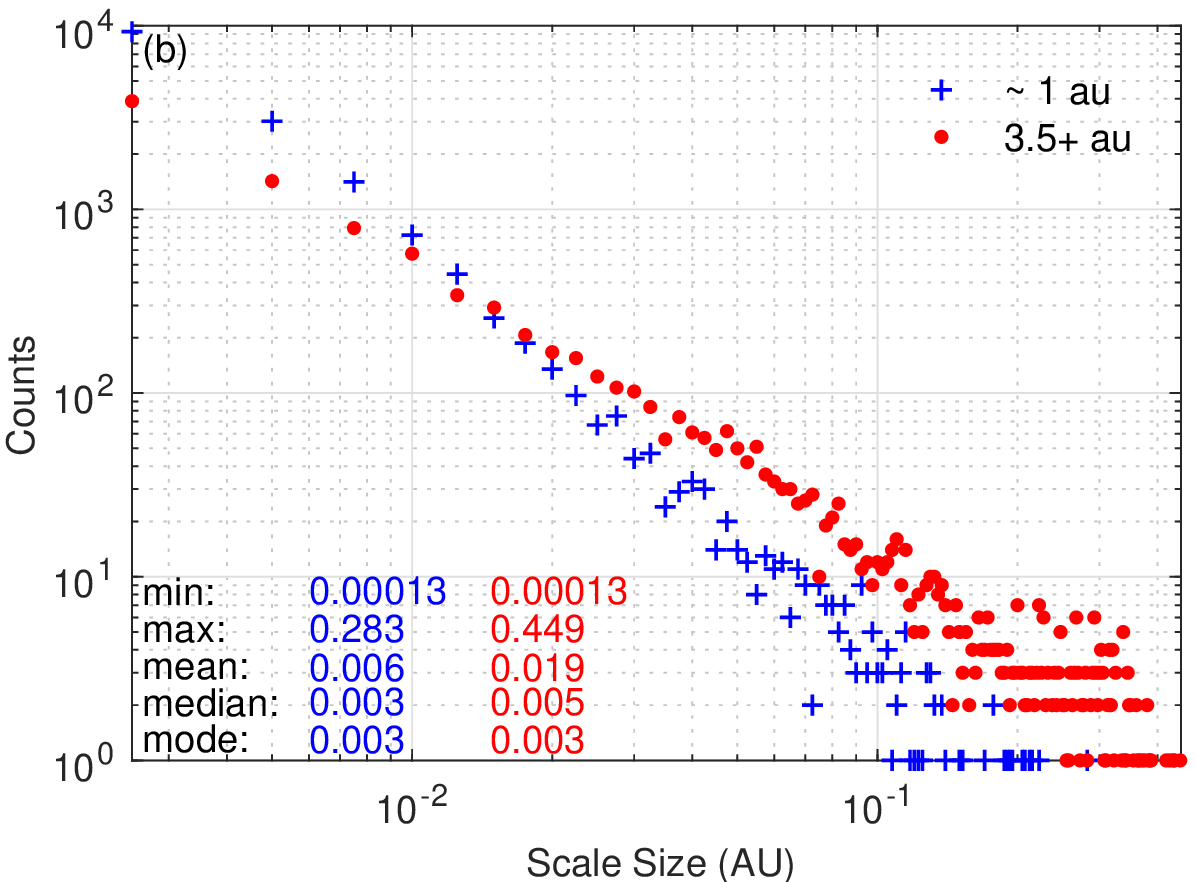}
\includegraphics[scale=0.66]{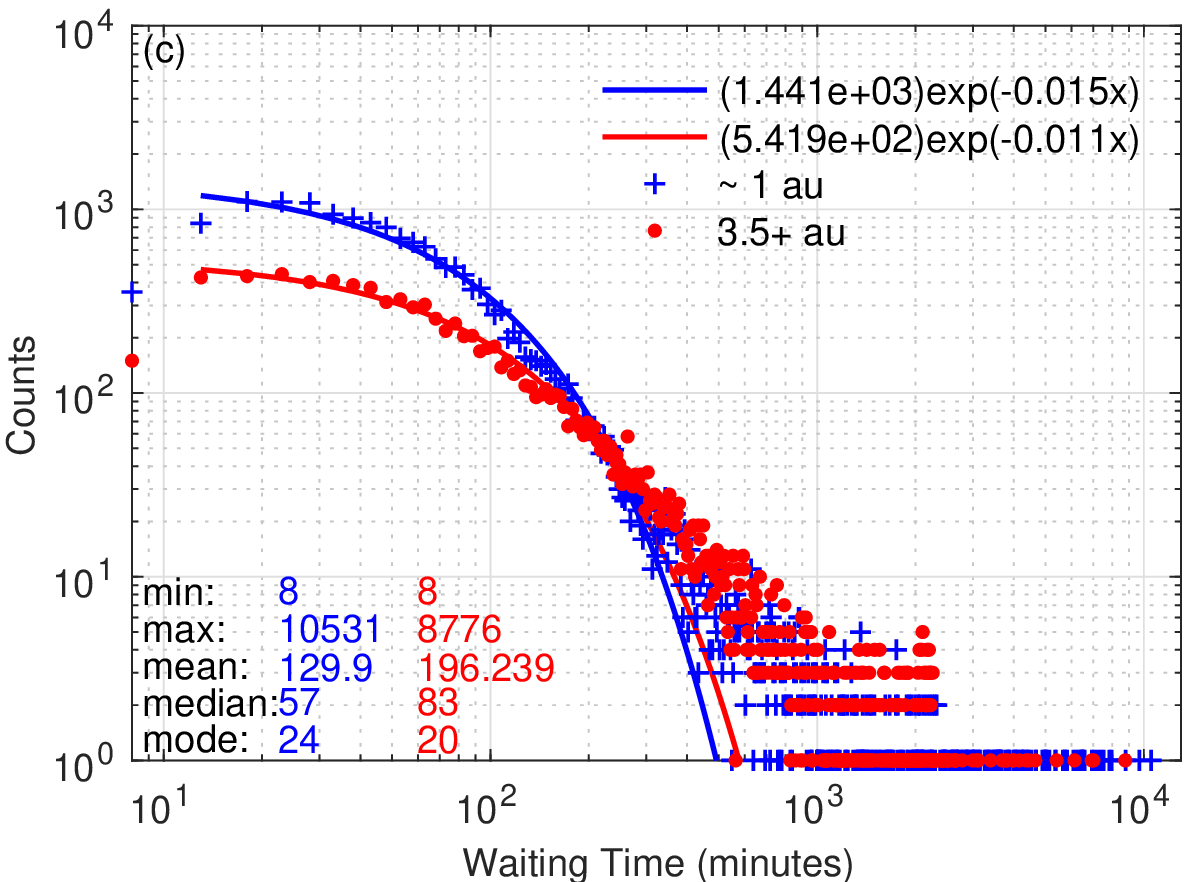}
\includegraphics[scale=0.66]{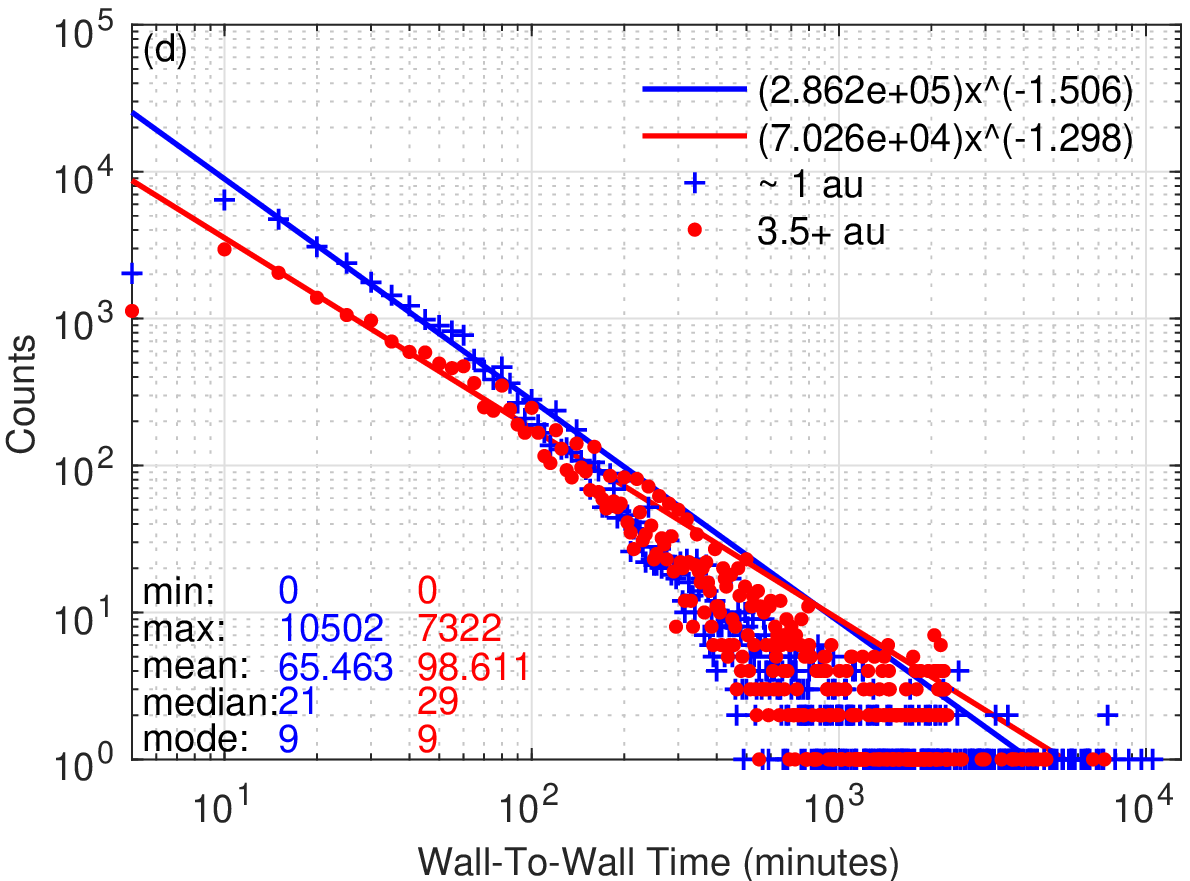}
\caption{Distributions of properties for the small-scale flux ropes at different radial distances: (a) Duration with 25 minutes as bin size. (b) Scale size with 0.0025 au as bin size. (c) The waiting time with 5 minutes as bin size. (d) The wall-to-wall time with 5 minutes as bin size. The blue crosses represent results from the \emph{ACE} spacecraft whereas the red dots are from the \emph{Ulysses} spacecraft. Exponential fitting curves and power-law function fitting curves are drawn in (c) and (d), respectively, with the corresponding fitting parameters denoted.}\label{fig:aceulypara1}
\end{centering}
\end{figure}

Figure \ref{fig:aceulypara1} is the set of distributions with selected fitting curves for \emph{ACE} and \emph{Ulysses} searching results. Figure \ref{fig:aceulypara1}a shows the distributions of duration. As suggested in \cite{Zheng2018,Hu2018}, events of small duration contribute the most to the occurrence rate of flux ropes. Both distributions exhibit power-law behavior, albeit with different power-law indices. In fact, although the maximum of \emph{ACE} flux rope duration is nearly the same as that of \emph{Ulysses}, the mean value is still less than the corresponding \emph{Ulysses} result (40 and 126 minutes, respectively). Also, flux ropes at far distances yield larger scale sizes as indicated in Figure \ref{fig:aceulypara1}b. The mean value is 0.019 au at larger distances whereas the mean of 1 au result is 0.006 au, which is smaller than the one detected by \cite{Moldwin1995}, i.e., 0.05 au. Moreover, Figure \ref{fig:aceulypara1}c shows that flux ropes at far distances tend to have longer average waiting time than they do at 1 au. On one hand, the exponential fitting curves are suitable for shorter waiting time. On the other hand, the power law fitting curves perform better for longer waiting time distributions. Figure \ref{fig:aceulypara1}d presents the distributions of wall-to-wall time. The power law fitting curves are shown for the smaller value portions of both distributions. The blue line has a break point near 80 minutes while the red line seems to have a break point beyond 100 minutes.

\begin{figure}
\begin{centering}
\includegraphics[scale=0.66]{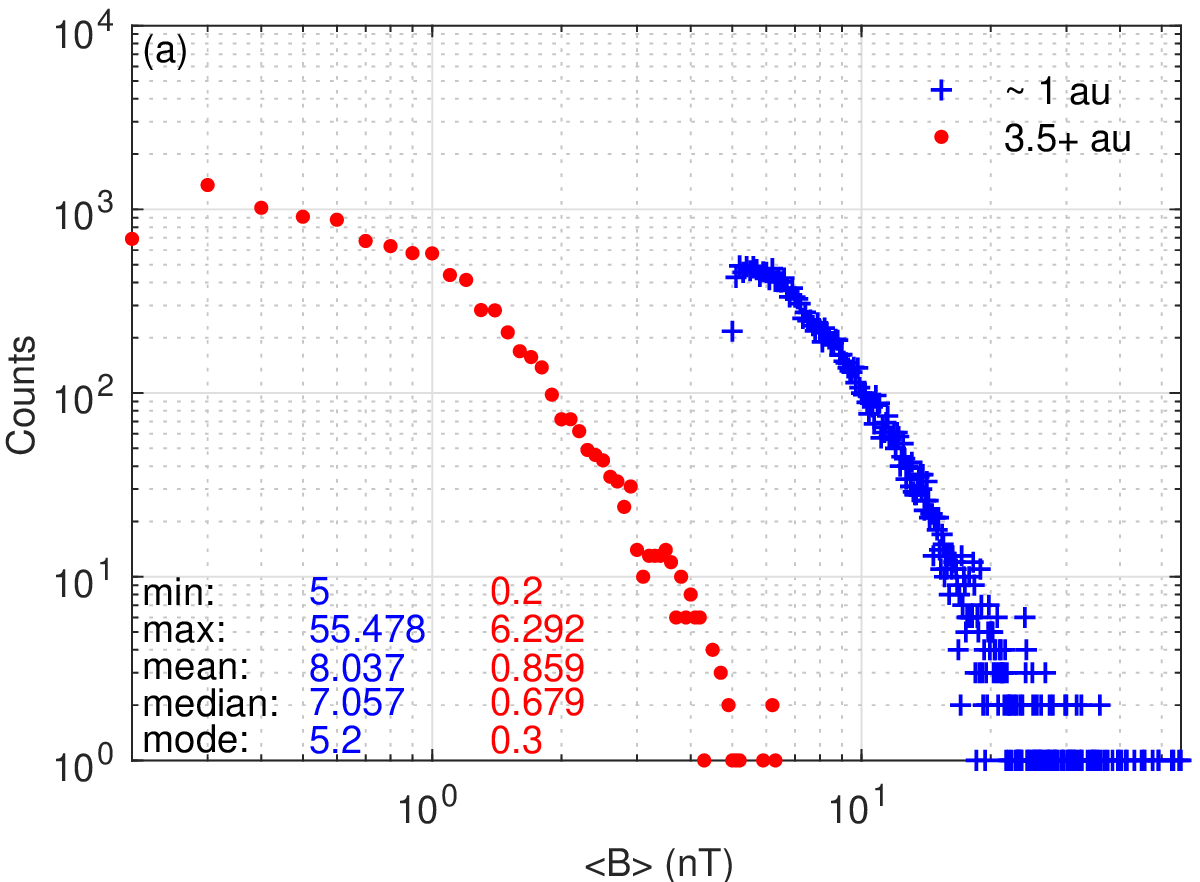}
\includegraphics[scale=0.66]{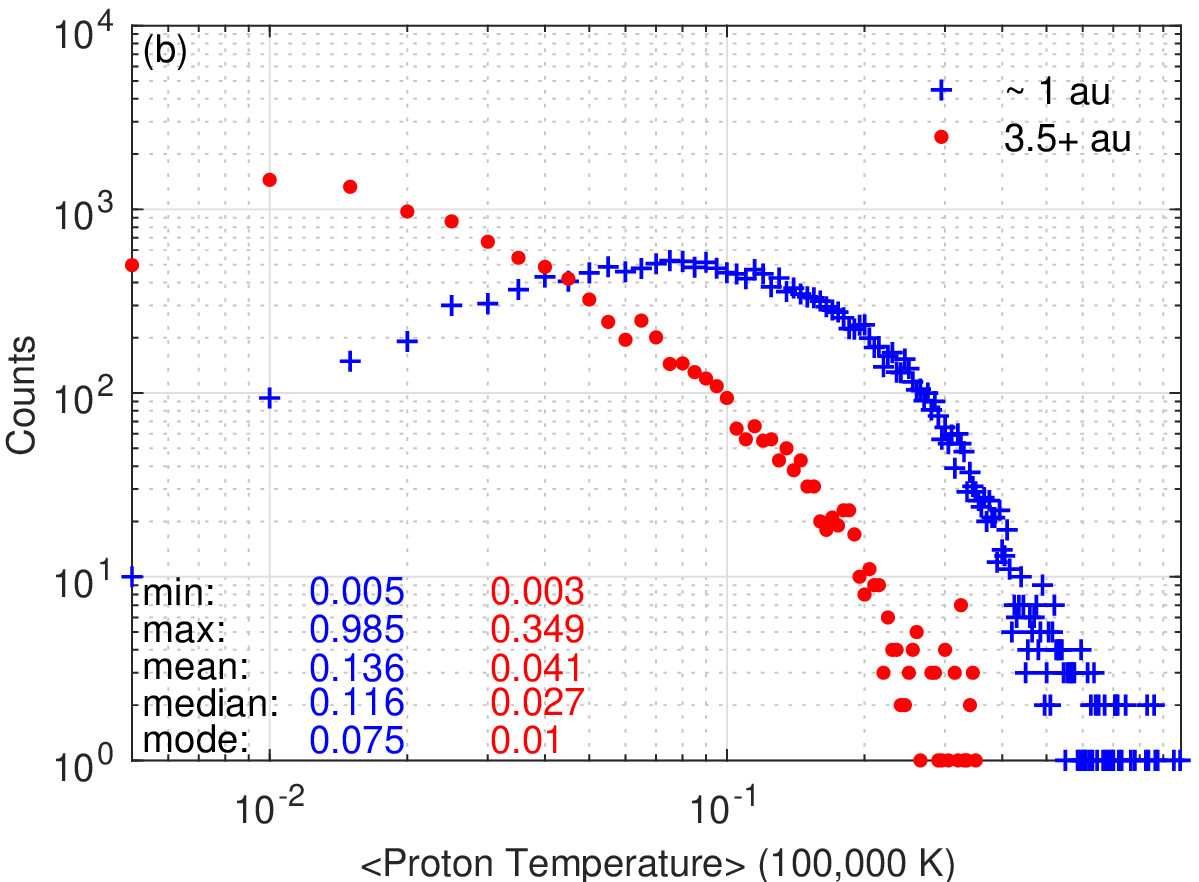}
\includegraphics[scale=0.66]{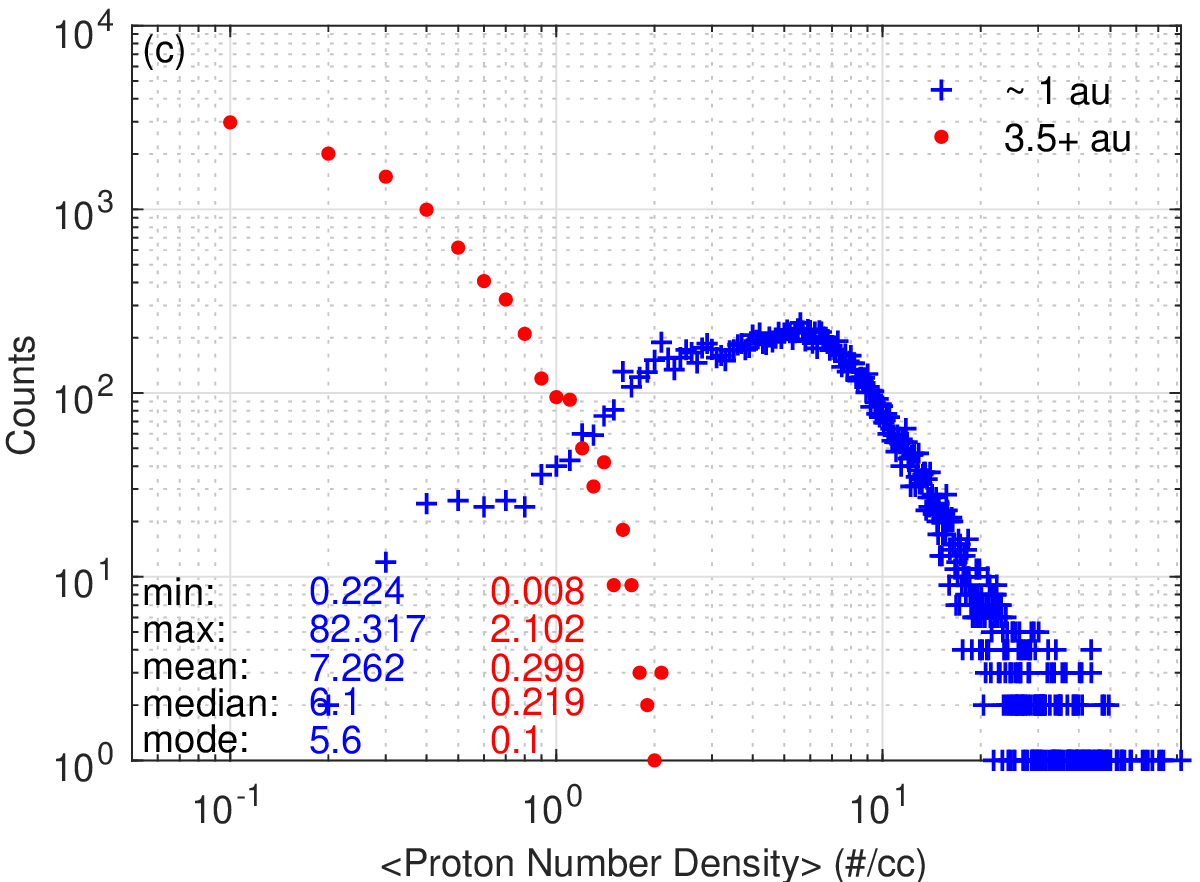}
\includegraphics[scale=0.66]{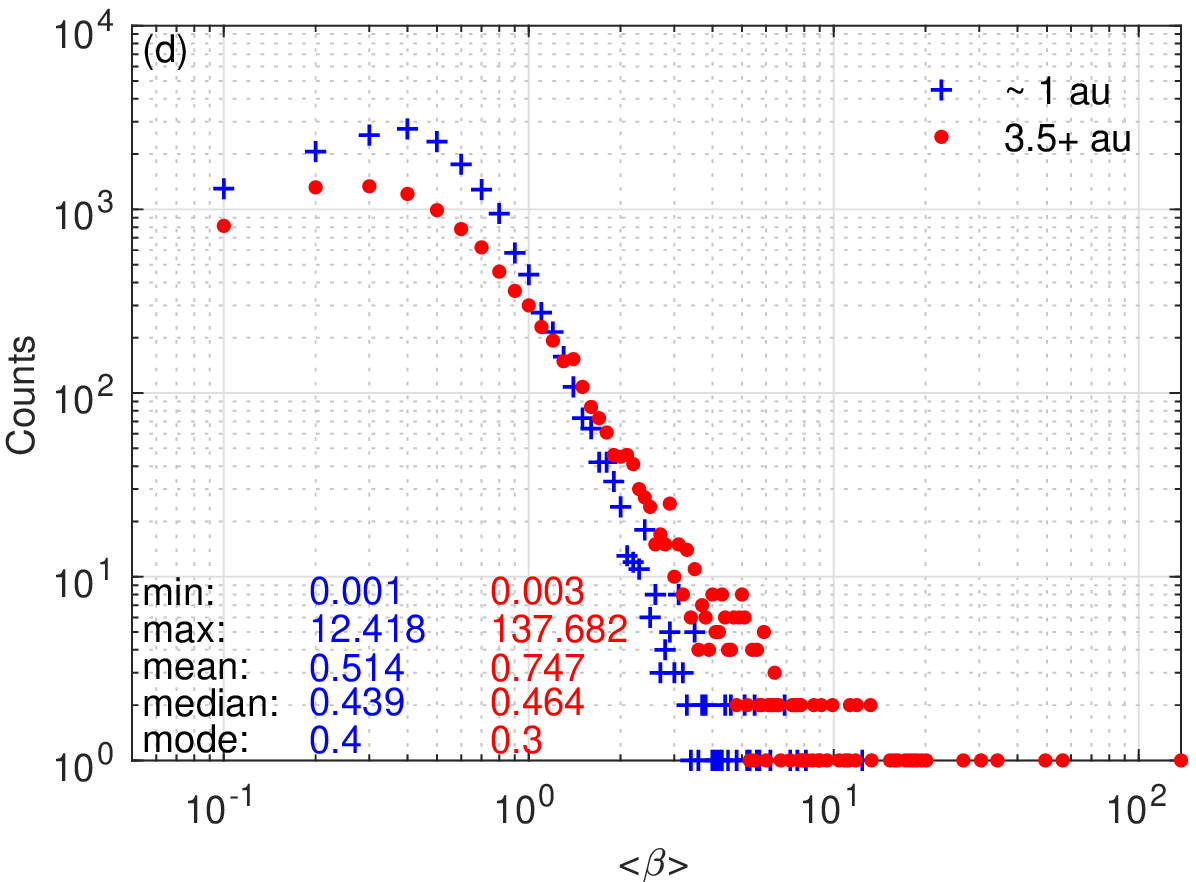}
\caption{Distributions of small-scale flux rope properties at different radial distances: (a) Average of the magnetic field magnitude with 0.1 nT as bin size. (b) The proton temperature $T_p$, 0.005 $\times$ $10^6$ K as bin size. (c) The proton number density, 0.1 $cm^{-3}$ as bin size. (d) The average proton $\beta$, 0.1 as bin size. The format is the same as Figure \ref{fig:aceulypara1}.}\label{fig:aceulypara2}
\end{centering}
\end{figure}

Figure \ref{fig:aceulypara2} demonstrates the rest of the of flux rope properties. Because all the results are within 30$^{\circ}$ latitude for \emph{Ulysses}, the radial distance becomes the primary factor for generating any differences. The magnetic field at 1 au is almost ten times larger than that at distances greater than 3.5 au. Figure \ref{fig:aceulypara2}b and c show that the distributions of proton temperature and number density also differ significantly due to the separation in radial distances. Combining all three parameters, the plasma $\beta$ (Figure \ref{fig:aceulypara2}d) at different radial distances exhibits distributions close to each other.

\section{Features under Different Solar Activity Levels}\label{sec:solar}
As discussed in Sections \ref{sec:overview} and \ref{sec:rad}, the database of SFRs at 1 au on the basis of \emph{Wind} and \emph{ACE} spacecraft measurements reveals that the occurrence of flux ropes has a solar cycle dependency and varies with the level of solar activity with a short time lag \citep{Zheng2018, Hu2018}. However, this cycle dependency for the \emph{Ulysses} monthly occurrence count seems to be modulated by the strong effects from the changing radial distance and latitudes. Such variations in occurrence rate are further diminished when a more strict Wal\'en slope threshold is applied to exclude possible Alfv\'enic structures or waves.

In order to further investigate whether different solar activity levels would have effects on features of small-scale flux ropes, we isolate three periods as marked in Figure \ref{fig:ulycount} during which \emph{Ulysses} traveled from one pole to the other over the largest latitudinal span. Fortunately, all these time periods are corresponding with either solar maximum or minimum periods as indicated by the sunspot numbers. Not only are the latitudes similar (the first two are from -80.2$^{\circ}$ to 80.2$^{\circ}$, and the last one is from -79.7$^{\circ}$ to 79.7$^{\circ}$), but also the radial distances are within the same range 1.34 $\sim$ 2.41 au. As there is almost no difference between the two minima, we combine these two periods and classify all three fast latitudinal scans into two groups: the solar minimum period which contains 2,912 flux ropes and solar maximum period which has 1,238 flux ropes, as indicated in Table \ref{table:table2}.

\begin{figure}
\begin{centering}
\includegraphics[scale=0.76]{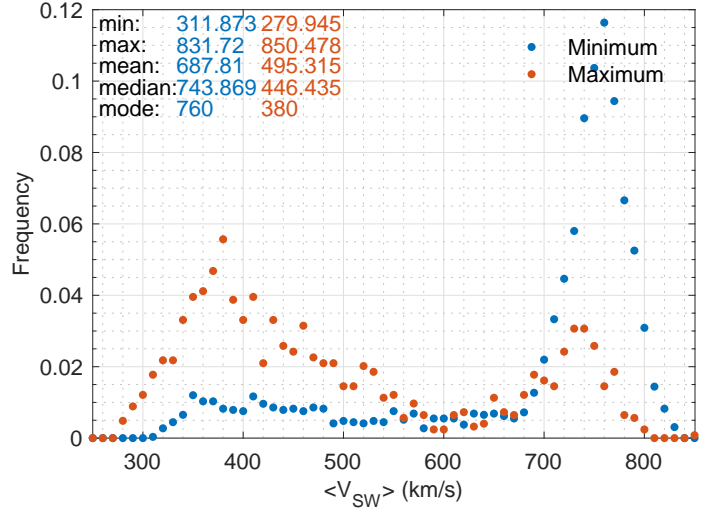}
\caption{Histogram of average solar wind speed under different solar activity levels. Flux ropes occurring at solar minima are denoted by blue dots while those at solar maxima are marked by dark orange dots for the time periods represented in Figure \ref{fig:ulycount}.}\label{fig:maximumvsw}
\end{centering}
\end{figure}

Figure \ref{fig:maximumvsw} is the distribution of solar wind speed for these two groups. Average solar wind speed of flux ropes occurring during the first and the third fast latitude scans are denoted by blue dots, and the rest occurring during the second scan is marked by dark orange dots. This type of distribution is typical or corresponding to the well-known fact that solar wind is dispersed at all latitudes during the solar maximum. As implied by the mean and mode values, high speed wind is dominant again during the solar minimum.

\begin{figure}
\begin{centering}
\includegraphics[scale=0.66]{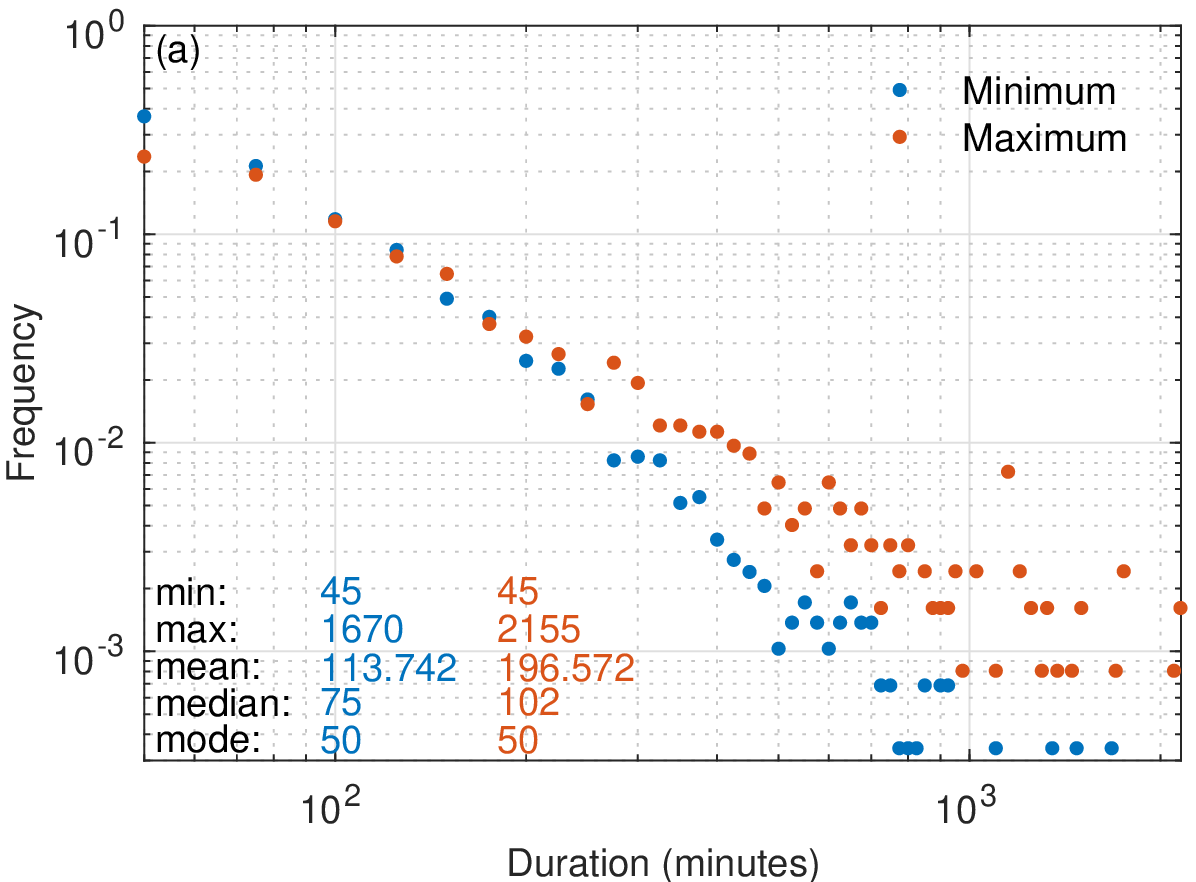}
\includegraphics[scale=0.66]{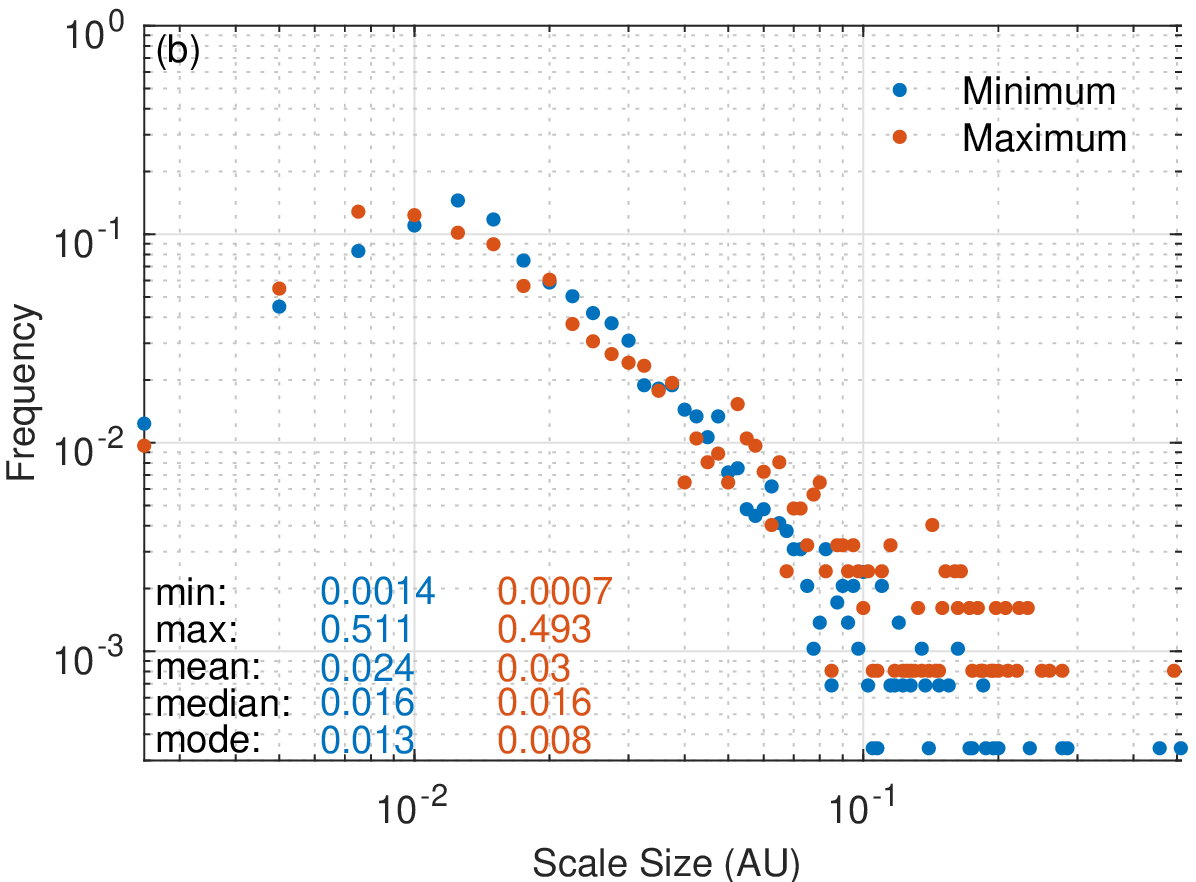}
\includegraphics[scale=0.66]{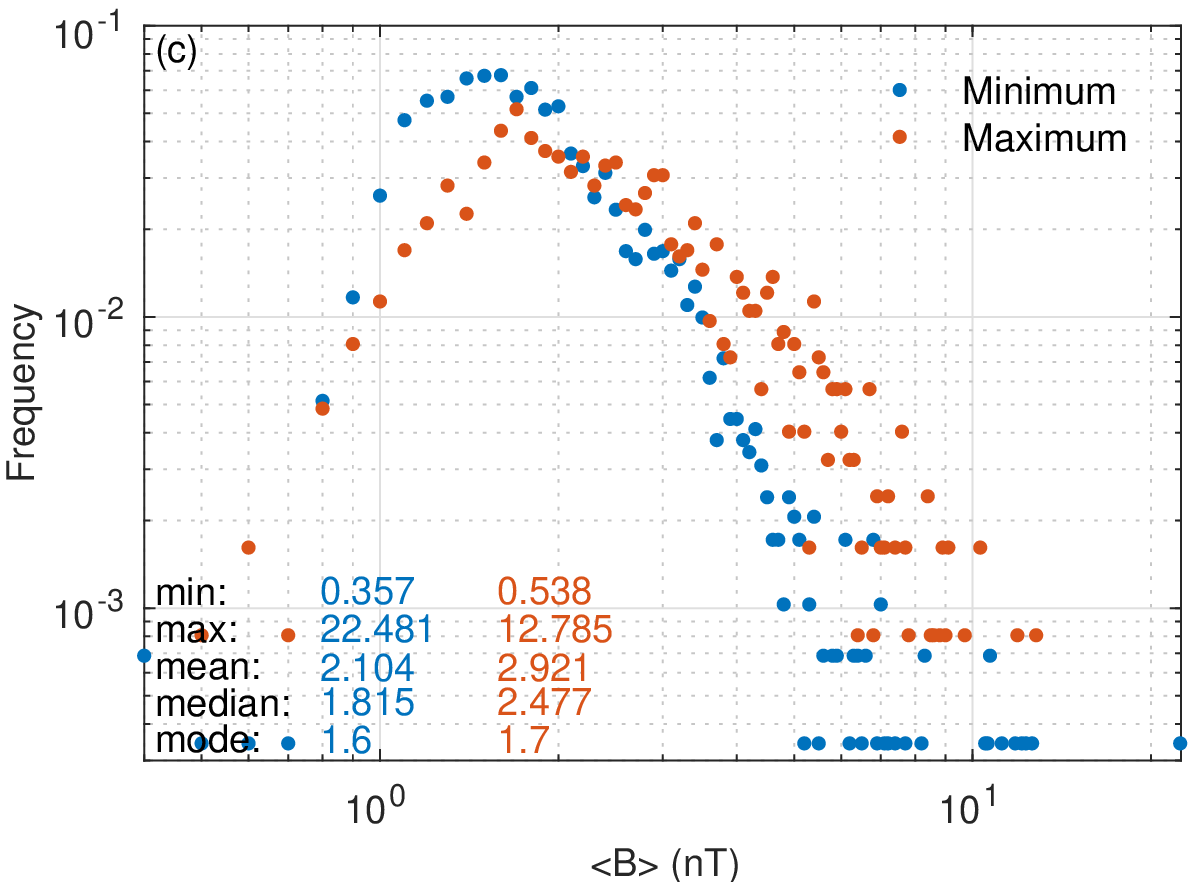}
\includegraphics[scale=0.66]{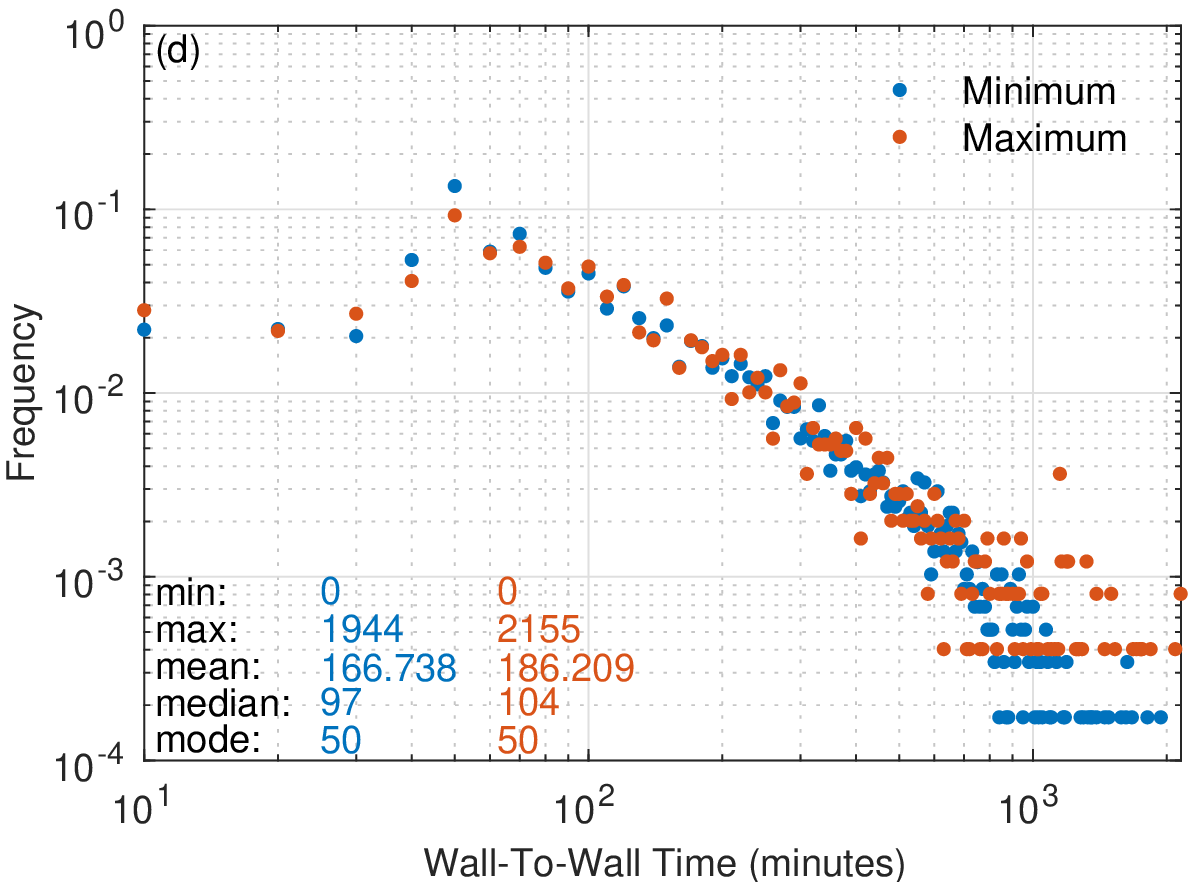}
\caption{Distributions of small-scale flux rope properties under different solar activities: (a) Duration. (b) Scale size. (c) Average magnetic field magnitude. (d) Distribution of wall-to-wall time. The format is the same as Figure \ref{fig:ulyparameter1}.}\label{fig:maxmin5}
\end{centering}
\end{figure}

The basic properties of flux ropes categorized by different solar activity levels are investigated, and Figure \ref{fig:maxmin5} shows selected plots of distributions. We caution, however, because of much reduced number of events in this category, we mostly limit our interpretations to the portions of the most abundant number of events, and refrain from any functional fittings. Figure \ref{fig:maxmin5}a, b, and d show that there is little difference between the two groups in the distributions of duration, scale size, and wall-to-wall time, especially for the small parameter values. Figure \ref{fig:maxmin5}c shows that the magnetic field is a little stronger during the solar maximum with the same range of radial distances and latitudes.

\section{Conclusions and Discussion}\label{sec:con}
We present the new database of 22,719 small-scale magnetic flux ropes detected from the in-situ \emph{Ulysses} measurements covering the entire mission. The approach is the newly developed automated detection algorithm based on the Grad-Shafranov reconstruction, which scans the entire time-series data through a sliding-window process, yielding the list of identified flux ropes with variable duration and a number of associated properties. These properties include the duration, the cross-section scale size, various average magnetic and plasma parameters for each flux rope interval, and additional derived parameters such as the waiting time and wall-to-wall time for the entire event list (database). We plan to make available the event lists on our designated database website, \url{http://www.fluxrope.info}. The characteristics of the SFRs are summarized and discussed in the following aspects: (1) how they depend on orbit latitude which uses 30$^{\circ}$ as the boundary separating the low and high latitude bands, (2) dependence on radial distance including 1 au detection results via \emph{ACE} in-situ measurements, and (3) variation under different solar activity levels during the \emph{Ulysses} fast latitudinal scans. All the properties are analyzed and interpreted in a statistical manner based on the above categorization (see also Table \ref{table:table2}). The main results are listed as follows.

\begin{enumerate}[1.]
\item The magnetic field and plasma parameters at different latitudes generally follow the distribution of solar wind speed. SFRs detected at high latitudes, where the high speed wind dominates, tend to have higher temperature ${T_p}$ and lower number density ${N_p}$, and vice versa. Moreover, due to the prevailing high speed wind at those latitudinal regions, SFRs tend to have shorter duration resulting in similar scale sizes when compared with regions around the ecliptic. The waiting time distribution exhibits exponential function behavior while the wall-to-wall time distribution has power law function behavior with a break point around 300 minutes. Most properties exhibit power-law distributions, but with different power-law indices especially for the SFRs at different radial distances.

\item Alfv\'enic structures or waves, and possible flux ropes with high Alfv\'enicity occur more often at high latitudes during solar minima. By applying a more strict Wal\'en slope threshold, the latitudinal effects on SFRs are reduced, but the radial distance effects remain. The distribution of duration, scale size and wall-to-wall time remain power law functions but with different power-law indices for the high latitude and low latitude groups.

\item Most of SFRs tend to lie on the local $RT$ plane and along the Parker spiral direction (i.e., to which the flux rope axis is parallel) near the ecliptic which is consistent with 1 au detection result.

\item For events within low latitudinal ranges, the distributions of solar wind speed at different radial distances are almost identical. The SFR magnetic field, however, and the plasma temperature and density vary substantially with increasing radial distances. Those flux ropes at far distances (greater than 3.5 au) are inclined to have larger scale sizes, longer duration and waiting time, possibly due to expansion or neighboring flux rope merging.

\item The 18.5 years' lifetime of the \emph{Ulysses} mission covered 1.5 solar cycles consisting of two solar minima and one solar maximum. Unlike the detection result at 1 au, the solar cycle dependency of SFRs via the \emph{Ulysses} measurements is modulated by the dependence on radial distances and helio-latitudes. When a more strict Wal\'en slope threshold ($\leqslant$ 0.1) is applied, the variations in the event occurrence rate are further suppressed.  In other words, the solar cycle dependency is diminished. Again, the main impact of the exclusion of possible Alfv\'enic structures or waves is at high latitudes where the number of SFRs has the most significant reduction due to a more strict Wal\'en slope threshold.

\item During three fast latitudinal scans, the flux ropes detected during the solar maximum have slightly stronger magnetic field. The other properties do not appear to differ significantly between the maximum and the minimum periods. Each period lasted for about a year only.

\end{enumerate}

One of the main applications of our database is to make connection to particle acceleration, which has been elucidated in both theory and simulation. Previous theoretical studies suggested that energetic ions can be accelerated when contracting or merging flux ropes exist \citep{Zank2014, Zank2015, leRoux2015a, leRoux2015b}. Recently, the theory has been extended in \cite{leRoux2018}. A set of equations for energetic particles is presented, which provides a self-consistent description of the energy exchange between suprathermal particles and SFRs for different SFR acceleration processes. Moreover, \cite{Zhao2018, Zhao2019} combined the theoretical prediction of particle acceleration with our detection result of SFRs from \emph{Ulysses} observations to show the unusual energetic particle flux enhancement as a result of SFR dynamics, for one particular time period of a few weeks. A number of SFRs was identified downstream of an interplanetary shock and in the neighboring corotation interaction region (CIR), together with the compressional waves associated with the CIR, as well as the heliospheric current sheet nearby. This complex multi-stream system provides a favored environment for the generation of SFRs. Thus, we expect to examine more observational cases which have particle acceleration signatures associated with SFRs and provide the additional observational evidence for the existing theories and simulation results in future study.

Furthermore, with the launch of the Parker Solar Probe which was designed to probe much closer to the Sun, we expect to have an extra detection point supplying our existing database with unprecedented data products to further investigate the origin and evolution of SFRs throughout the inner heliosphere. In particular, the effect due to varying radial distances and the possible radial evolution of these structures have yet to be further elucidated by examining additional in-situ spacecraft datasets.

\acknowledgments
We would like to thank the teams of the \emph{Ulysses}, \emph{Wind} and \emph{ACE} spacecraft missions as well as the Coordinate Data Analysis Web (CDAWeb) of NASA GSFC for available in-situ data. We acknowledge NASA grants NNX15AI65G, NNX17AB85G, 80NSSC18K0623, 80NSSC19K0276 and subawards SAO SV4-84017, and NSF grant AGS-1650854 for support.


\end{document}